%% file: main.tex
\PassOptionsToPackage{fleqn}{amsmath}
\documentclass{sig-alternate}
\usepackage{mathptmx} 

\newcommand{\ignore}[1]{}
\usepackage{fancyhdr}
\usepackage[normalem]{ulem}
\usepackage[hyphens]{url}
\usepackage{microtype}

\usepackage{amsfonts}

\usepackage{flushend}
\usepackage{comment}
\usepackage{amsmath}
\usepackage{mathtools}
\usepackage{wrapfig,lipsum,booktabs}
\usepackage{multirow}
\usepackage{color}

\usepackage{rotating}
\usepackage[flushleft]{threeparttable}
\usepackage{multicol}

\usepackage{algorithm}
\usepackage{algorithmic}
\usepackage{soul}

\usepackage{cite}

\newbool{inccomment}
\setbool{inccomment}{true}
\newcommand{\lhs}[1]{\ifbool{inccomment}{{\color{red}LHS: #1}}{}}

\usepackage{float}

\newcommand{\hpcasubmissionnumber}{18}

\fancypagestyle{firstpage}{
  \fancyhf{}
\setlength{\headheight}{50pt}

  \fancyhead[C]{\normalsize{HPCA 2019 Submission
      \textbf{\#\hpcasubmissionnumber} -- Confidential Draft -- Do NOT Distribute!!}}
  \pagenumbering{arabic}
}

\newcommand{\hypar}{\textsc{HyPar}}

\title{HyPar: Towards Hybrid Parallelism
 for \\
Deep Learning Accelerator Array}
\author{\Large  Linghao Song$^\dagger$, Jiachen Mao$^\dagger$, Youwei Zhuo$^\ddagger$, Xuehai Qian$^\ddagger$, Hai Li$^\dagger$, Yiran Chen$^\dagger$ \vspace{3pt}\\ 
\it \Large $^\dagger$Duke University, $^\ddagger$University of Southern California \vspace{6pt}\\
\large
\{linghao.song, jiachen.mao, hai.li, yiran.chen\}@duke.edu,
\{youweizh, xuehai.qian\}@usc.edu}

\begin{document}

\maketitle
\pagestyle{plain}


\begin{abstract}
\input{sec_abs}

\end{abstract}

\section{Introduction}
\label{sec_intro}
\input{sec_intro}

\section{Background and Motivation}
\label{sec_back}
\input{sec_back}
\section{Communication Model}
\label{sec_commodel}
\input{sec_commodel}

\section{Layer Partition}
\label{sec_partition}

\input{sec_partition}

\section{hypar Architecture}
\label{sec_implementation}
\input{sec_implementation}

\section{Evaluation}
\label{sec_eval}

\input{sec_eval}

\section{Conclusion}
\label{sec_conc}
\input{sec_conc}


\bibliographystyle{ieeetr}
\bibliography{main}

\end{document}

%% file: sec_abs.tex
With the rise of artificial intelligence in recent years, Deep Neural Networks (DNNs) have been widely used
in many domains. 
To achieve high performance and energy efficiency,
hardware acceleration (especially inference) of DNNs is intensively studied
both in academia and industry.
However, we still face two challenges:
large DNN models and datasets, which incur frequent off-chip memory accesses; and 
the training of DNNs, which is not well-explored in recent accelerator designs.
To truly provide high throughput and energy efficient acceleration for the training of deep and large models, 
we inevitably need to use multiple accelerators to explore 
the coarse-grain parallelism, compared to the fine-grain parallelism
inside a layer considered in most of the existing architectures.
It poses the key research question to seek the best
organization of computation and 
dataflow among accelerators.

In this paper, inspired by 
recent work in machine learning systems,
we propose a solution \hypar\ to determine layer-wise parallelism for deep neural 
network training with an array of DNN accelerators. 
\hypar\ partitions the feature map tensors (input and output), the kernel tensors, 
the gradient tensors, and the error tensors for the DNN accelerators. 
A partition constitutes the choice of parallelism for weighted layers. 
The optimization target is to search a partition that
minimizes the total communication during training a complete DNN.
To solve this problem, 
a communication model is used
to explain the source and amount of 
communications. 
Then, we use a hierarchical layer-wise dynamic programming
method to search for the partition for each layer. 
\hypar\ is practical: the time complexity for the partition 
search in \hypar\ is {\em linear}.
We apply this method in 
an HMC-based DNN training architecture to minimize data movement.
We evaluate \hypar\ with ten DNN models from classic Lenet to large-size model VGGs, and the number of weighted layers of these models range from four to nineteen. Our evaluation finds that: 
the default Model Parallelism is indeed the worst;
the default Data Parallelism is not the best;
but hybrid parallelism can be better than either the default Data Parallelism or Model Parallelism in DNN 
training with an array of accelerators.  Our evaluation shows that \hypar\ achieves a performance gain of 3.39$\times$ and an energy efficiency gain of 1.51$\times$ compared to Data Parallelism on average, and \hypar\ performs up to 2.40$\times$ better than ``one weird trick''.

%% file: sec_intro.tex
With the rise of artificial intelligence in recent years, Deep Neural Networks (DNNs) have been widely used because of their high accuracy, excellent scalability, and self-adaptiveness properties \cite{goodfellow2016deep, bengio2015deep}. Many applications employ DNNs as the core technology, such as face detection \cite{sun2015deepid3}, speech recognition \cite{hinton2012deep}, scene parsing~\cite{farabet2013}.

DNNs are computation and memory intensive and pose intensive challenges to the 
conventional Von Neumann architecture where 
computation and data storage are separated.
For example, AlexNet~\cite{alexnet2012} performs 
$10^9$ operations in processing just one image data.
During processing, a large amount of data movements are incurred 
due to the large number of layers and 
millions of weights. 
Such data movements quickly become a performance bottleneck due to 
limited memory bandwidth and more importantly, an energy bottleneck.
A recent study~\cite{farmahini2015} showed that data 
movements between CPUs and off-chip memory 
consumes two orders of magnitude more energy than a floating point operations.
Clearly, the cost of computation and data movement are serious 
challenges for DNNs.

To achieve high performance and energy efficiency,
hardware acceleration of DNNs is intensively studied
both in academia~\cite{liu2016cambricon,zhang2016cambricon,chen2014diannao,chen2014dadiannao,du2015shidiannao,liu2015pudiannao,chen2016eyeriss,chen2017eyeriss,chen2017using,han2016eie,du2015neuromorphic,kim2016neurocube,lu2017flexflow,ren2017sc,gao2017tetris,furber2013overview,zhang2015optimizing,qiu2016going,motamedi2016design,suda2016throughput,zhang2016caffeine,han2017ese,ma2017optimizing,xu2017parallel,zhang2017improving,esmaeilzadeh2012neural,farabet2011neuflow,venkataramani2017scaledeep,ji2016neutrams,na2016speeding,parashar2017scnn,alwani2016fused, shen2016overcoming, shen2017escher,shen2017maximizing,mirhoseini2016perform,razlighi2017looknn,takhirov2016energy,ko2017design,sharma2016high,reagen2016minerva,zhang2017frequency,bing2019red,chen2018regan,chen2018emat,li2018reram,ji2018recom,qiao2018atomlayer,mao2017modnn,mao2017mednn,mao2017adalearner,tang2017binary,jiang2017xnor,deng2018dracc,lou20183dict,ji2019HUBPA,liu2017mt,liu2018pt,zhang2018dnnbuilder,wang2018snrram,li2017drisa,leestitch,liu2018processing,yang2018quantized,liu2015reno,yan2018celia,wang2017group,albericio2016cnvlutin,judd2016stripes,mahajan2016tabla,bojnordi2016memristive,yu2017scalpel,albericio2017bit,ding2017c,park2017scale,cai2018vibnn,Yazdanbakhsh2018gan,Akhlaghi2018SnaPEA,Hegde2018ucnn,park2018energy,song2018prediction,sharma2018bit,deng2018perm} and industry~\cite{tpu,tpu2017,merolla2014million,esser2015backpropagation,esser2016convolutional,intel,intel_loihi,nvidia,qualcomm,deephi,microsoft_acc}.
In particular, several major companies developed 
{\em 1)} DNN accelerators, e.g., Google TPU \cite{tpu,tpu2017}, and neuro-processors, e.g., IBM TrueNorth \cite{merolla2014million,esser2015backpropagation,esser2016convolutional}; 
{\em 2)} corresponding standards, architectures, and platforms \cite{intel,nvidia,qualcomm,deephi}.
In academia, 
Eyeriss \cite{chen2017eyeriss} is a representative design of spatial architecture to coordinate dataflow between processing engines (PEs). 
Neurocube \cite{kim2016neurocube} takes the advantage of in-memory processing by deploying PEs in hybrid memory cubes (HMCs)~\cite{hmcspec} with a programmable data packet scheme. 
Flexflow \cite{lu2017flexflow} is a systolic architecture with tiling optimization.
Furthermore, inter-layer data flow for DNN acceleration are 
considered in~\cite{alwani2016fused, shen2017maximizing,song2017pipelayer}. 

Despite the explosion of neural network
accelerators, there are still fundamental 
challenges.
First, with large model size (e.g., for ImageNet dataset \cite{deng2009imagenet}), accelerators suffer from the frequent access to off-chip DRAM, which consume significant energy, e.g., 200$\times$ compared to on-chip SRAM~\cite{chen2017eyeriss,han2016eie}, and can thus 
easily dominate the whole system power consumption.
Second, almost all of the recently proposed DNN accelerators only focuses on DNN inference.


The \emph{inference} of deep neural networks is a forward progress of input images 
from the first layer to the last layer.
Kernels (weights) of a network are obtained through training
before the inference.
The computations are performed one layer after another.
The \emph{training} of deep neural networks is more complex than inference 
due to more computations and additional data dependencies. 
Besides data forward, error backward and gradient computation
are two additional computation steps to generate new weights in training.
With the increasing importance of deep learning, 
we argue that DNN training acceleration is 
a crucial problem.
Currently, DNNs are typically trained by high-performance computer systems 
with high-end CPUs/GPUs, which are not performance and energy efficient.
While many accelerators focused on acceleration for DNN inference, 
training was only considered in a few existing accelerators \cite{kim2016neurocube, song2017pipelayer} in restricted manner.

To truly provide high throughput and energy efficient acceleration for training deep and large models \cite{simonyan2014very, he2016deep}, 
we inevitably need to use {\em multiple accelerators} as a general architecture to explore 
the coarse-grain parallelism, compared to the fine-grain parallelism
inside a layer considered in the existing research \cite{chen2017eyeriss, kim2016neurocube, lu2017flexflow,tpu2017}. 
It poses the key research question to seek the best
organization of computation and 
dataflow among accelerators.

The problem is challenging due to the complex interactions
between the type of parallelism and different layers. Assume we have $N$ accelerators,
in {\em data parallelism}, a batch of data is partitioned into $N$ parts, 
while the model (weights) are duplicated $N$ times. Each accelerator
holds one part of the partitioned data and a complete copy of the model. 
It incurs no communication in data forward and error backward.
To update the weights, 
each accelerator requires remote access to the gradient in the other accelerators, thereby results in communication. 
In {\em model parallelism}, the kernel is partitioned into $N$ parts, and feature maps are partitioned accordingly.
Opposite to data parallelism, 
it incurs communications in data forward but 
no communication in error backward and weight updating. 
More details are discussed in Section~\ref{par_basic}.

The current accelerator designs {\em do not provide a good answer}
on how to determine parallelism for multiple accelerators, because they focus on the acceleration of 
intra-layer (fine-grain parallelism) computation and assume that the needed data are already 
in memory.
Clearly, the solution is nontrivial.
We cannot simply partition all layers in data parallelism or
model parallelism manner, because different layer types 
imply different choices. 
Moreover, networks also have various connection between different layers. 
Overall, it can be phrased as a complex optimization problem.

Recently, mixed model and data parallelism is explored in
deep learning accelerator architectures \cite{song2017pipelayer,kim2016neurocube}
multi-GPU training systems~\cite{krizhevsky2014one,tofu,tofu2}.
Inspired by ~\cite{tofu,tofu2}, this paper
leverages both model and data parallelism
in {\em each layer} to minimize 
communication between accelerators. 
Specifically, we propose 
a solution \hypar\ 
to determine layer-wise parallelism for deep neural network training with an array of DNN accelerators. 
The goal of \hypar\ is to 
partition the feature map tensors (input and output), the kernel tensors, 
the gradient tensors, and the error tensors among the DNN accelerators. 
A partition is determined by 
the choice of parallelism for all weighted layers. 
The optimization target is to search a partition that
minimizes the total amount of communication during training a complete deep neural network.
In \hypar, a {\em communication model} 
 is proposed to explain where
the communication comes from in partitioned tensors (of feature maps and kernels), and determine the amount of communication.
To search for the partition for 
each layer, 
we apply the dynamic programming-based
search and recursive partitioning method proposed in \cite{tofu,tofu2} which partitions
on tensors and feature maps, respectively.
We apply the similar approach to search the 
best parallelism configuration among 
accelerators on a set of coaleased tensors.
We show that \hypar\ is practical: the time complexity for the partition 
search in \hypar\ is {\em linear}.
We apply this method in 
an HMC-based DNN training
architecture with an array of sixteen accelerators 
organized in four hierarchical levels 
and minimize the data movement.

Based on the architecture, 
we evaluate \hypar\ with ten DNN models from classic Lenet to large-size model VGGs, and the number of weighted layers of these models ranges from four to nineteen. Our evaluation shows that:
the default model parallelism is indeed worst;
the default data parallelism is also not the best;
but hybrid parallelism can be better than either the default data parallelism or model parallelism in DNN 
training with an array of accelerators.  Our evaluation shows that \hypar\ achieves a performance gain of 3.39$\times$ and a energy efficiency gain of 1.51$\times$ compared to data parallelism on average.
In addition, we also study the scalability and
the effects of network topology to 
provide deeper understanding of the \hypar\ architecture.

This paper is organized as follows. 
Section \ref{sec_back} introduces DNN background, parallelism in DNN computation, 
DNN accelerators and our motivations. 
Section \ref{sec_commodel} proposes a communication model to quantify the communication in DNN computation.
Section \ref{sec_partition} proposes the partition algorithm based on the communication model 
to determines the parallelism for each layer in training.
Section \ref{sec_implementation} presents 
an HMC-based accelerator array in which the partitions are generated by \hypar.
Section \ref{sec_eval} evaluates the performance, energy efficiency and communication of
\hypar\ architecture in DNN training, and conducts studies to provide further insights. Section \ref{sec_conc} concludes the paper.

%% file: sec_back.tex
\subsection{Inference and Training of DNNs}
\label{dnn_basic}

The \emph{inference} of deep neural networks is a forward progress of input data (typically images) from the first layer to the last layer.
Kernels (weights) of a network are obtained through training
before the inference. 
Images are typically grouped into batches to ensure high throughput with efficient forward propagation. 
The computations are performed one layer after another.
For a convolutional layer $l$, we use $\mathbf{F}_{l}$ to represent feature maps of this layer, and use $B$ to denote the batch size. 
We assume that each feature map is a three-dimensional tensor, 
with a height of $H$, a weight of $W$ and a depth of $C_l$ ($C_l$ is also the number of channels of Layer $l$).
The size of the feature map slice is $[H_l\times W_l\times C_l]$. 
Thus, $\mathbf{F}_{l}$ is of size $B\times[H_l\times W_l\times C_l]$. 
The kernel $\mathbf{W}_{l}$ has a size of $[K\times K\times C_l]\times C_{l+1}$, where $K$ is the height/width of kernels and $C_{l+1}$ is the number of channels of next layer, Layer $l+1$. 
$f(\cdot)$ is an activation function, performing element-wise non-linear operations. We use $\otimes$ to denote convolutions. The inference (forward propagation) can be represented as,
{
\vspace{0.0cm}
\setlength{\mathindent}{2.6cm}
\begin{equation}
\label{eq_forward}
\mathbf{F}_{l+1} = f\left(\mathbf{F}_{l}\otimes\mathbf{W}_{l} \right)
\end{equation}
\vspace{-0.4cm}
}

The \emph{training} of deep neural networks is more complex than inference. 
The purpose of training is to tune the kernels to reduce the value of a loss function. The loss function computes the difference between the output of a neural network with the ground truth (i.e., labels) of a group of input images. L2-norm and softmax are two examples of loss functions. 
Besides forward, \emph{error backward} and \emph{gradient computation} are two additional computation steps to generate new weights in training. We use $\mathbf{E}_{l}$ to represent errors for Layer $l$, the error backward can be represented as Equation \ref{eq_backpropagation}, 
{
\setlength{\mathindent}{2.3cm}
\begin{equation}
\label{eq_backpropagation}
\mathbf{E}_{l} = \left(\mathbf{E}_{l+1}\otimes\mathbf{W}^{*}_{l}\right)\odot
f^{\prime}\left(\mathbf{F}_{l} \right)
\end{equation}
}where $\mathbf{W}^{*}$ is a reordered form of $\mathbf{W}$ (if $\mathbf{W}$ is a matrix then $\mathbf{W}^{*} = \mathbf{W}^{\top}$), $\odot$ is an element-wise multiplication and $f'(\cdot)$ is the derivative of $f(\cdot)$.
The gradient computation is, 
{
\setlength{\mathindent}{2.8cm}
\begin{equation}
\label{eq_gradient}
\triangle\mathbf{W}_{l} = \mathbf{F}^{*}_{l}\otimes\mathbf{E}_{l+1}
\end{equation}
}with $\triangle\mathbf{W}_{l}$, we can update $\mathbf{W}_{l}$.

\subsection{Parallelisms}
\label{par_basic}

For DNNs training, the training data samples (images) are grouped into batches. For each epoch, a batch of data need to perform forward, 
error backward and gradient computation. 
Since training requires more computations, it is typically conducted with 
multiple DNN accelerators. 
In this context, parallelism needs to be considered among the accelerators.
{\em Data Parallelism} \cite{li2014scaling, li2014communication} and {\em Model Parallelism} \cite{coates2013deep, dean2012large} are the two types of parallelism used in DNN training.
In Data Parallelism, all accelerators hold a copy of model, but data (training samples) are partitioned into parts and each accelerator processes one part. In Model Parallelism, all accelerators process on the same data (training samples), but the whole model is partitioned and each accelerator holds a part of the model.

Whether to use Data Parallelism or 
Model Parallelism is currently 
determined empirically. 
For neural networks with rich convolutions, 
Data Parallelism is employed, while 
for neural networks with large model size, 
Model Parallelism is employed. 
Thus for the training of deep learning (which usually contains 
a lot of convolutions), 
Data Parallelism is the default setting 
\cite{li2014scaling, li2014communication}. 
Krizhevsky proposed ``one weird trick'' \cite{krizhevsky2014one} 
to outperform the default Data Parallelism, 
where convolutional layers are configured 
with data parallelism and fully connected layers
are configured with model parallelism. It is called ``weird'' because that method works but why is works was not known.


\subsection{DNN Accelerators}

Many DNN accelerators were proposed to optimize the data flow to 
improve performance and energy efficiency. Eyeriss \cite{chen2016eyeriss,chen2017eyeriss} is a representative design which employs a spatial data flow to share data between processing engines (PEs). Neurocube \cite{kim2016neurocube} takes the advantage of in-memory processing by deploying PEs in hybrid memory cubes (HMCs)~\cite{hmcspec} with a programming data packet scheme. Flexflow \cite{lu2017flexflow} is a systolic architecture with tiling optimization. 
MAESTRO \cite{kwon2018maestro} explored even five types 
of fine-grained on-chip data flow 
for DNN accelerator.
All of these accelerators 
focus on intra-layer computations in: computations for no more than one layer are performed at one time slot. They are all about the design of a stand-alone accelerator, which is orthogonal to this work.

In comparison, inter-layer data flow for DNN acceleration are 
considered in~\cite{alwani2016fused, shen2017maximizing,song2017pipelayer}. Alwani {\em et al.} \cite{alwani2016fused} proposed the fused-layer pipelining for DNN accelerators.
Shen {\em et al.} \cite{shen2017maximizing} further optimized the tiling parameters for inter-layers. 
Song {\em et al.} \cite{song2017pipelayer} 
proposed an inter-layer accelerator for DNN training.

\subsection{Motivation}
\label{motive}


With the increasing importance of deep learning, 
we argue that DNN training acceleration is a
crucial problem.
Currently, DNNs are typically trained by high-performance computer systems 
with high-end CPUs/GPUs, which is not computation and energy efficient.
For this reason, DNN training acceleration is of high 
interest to the companies with huge amount of data. 
For example, Google released a new version of TPU \cite{TPU-2} for training after a first version of TPU \cite{tpu2017} designed for inference.  
In research community, 
many accelerators focused on accelerating
DNN inference, and
training was only considered in a few existing accelerators \cite{kim2016neurocube, song2017pipelayer} in restricted manner.  
Neurocube \cite{kim2016neurocube} partitions model into HMC vaults, but 
does not consider the parallelism between HMCs. 
Among vaults, it assumes fixed parallelism setting
for all layers, 
which may not be the best for all networks. 
With inter-layer design (model parallelism), Pipelayer \cite{song2017pipelayer} 
performs the computations of different layers simultaneously
in different processing units of the accelerator. Pipelayer also
used a intra-layer parallelism, which is actually intra-layer data
parallelism, to boost performance. However, the details of data
movement for intra-layer and inter-layer parallelism was yet to be explored.


To truly support high throughput and energy efficient training
acceleration of deeper and large models \cite{simonyan2014very, he2016deep}, 
we eventually need to use {\em multiple accelerators} to explore 
the coarse-grain parallelism, compared to the fine-grain parallelism
inside a layer. It requires a systematic study to seek the best
organization of compute and 
dataflow among an array of accelerators.

This problem is {\em unsolved} by existing solutions \cite{chen2014diannao, chen2014dadiannao, chen2017eyeriss, lu2017flexflow}.
They only consider the acceleration of 
intra-layer computation and assumes that the needed data are already 
in memory. For a stand-alone accelerator that processes a layer separately, 
it is an acceptable assumption as the focus is the fine-grained 
computation inside the layer. With an array of accelerators,
the input data of an accelerator (for the 
current layer) are
potentially produced by other 
accelerators (for the previous layer), 
the computation and dataflow organization affect the {\em data 
movement}, which is a critical factor affecting the performance. 
The recent 
research efforts on mixed parallelism \cite{krizhevsky2014one,song2017pipelayer,tofu}, and layer-wise parallelism configuration and
the partitioning to minimizing communication
motivate us to develop a solution to 
determine tensor partition and dataflow organizations among layers for 
neural network training with an array of DNN accelerators.

%% file: sec_commodel.tex
\begin{figure*}[tb]
\centering
\includegraphics[width=1.7\columnwidth]{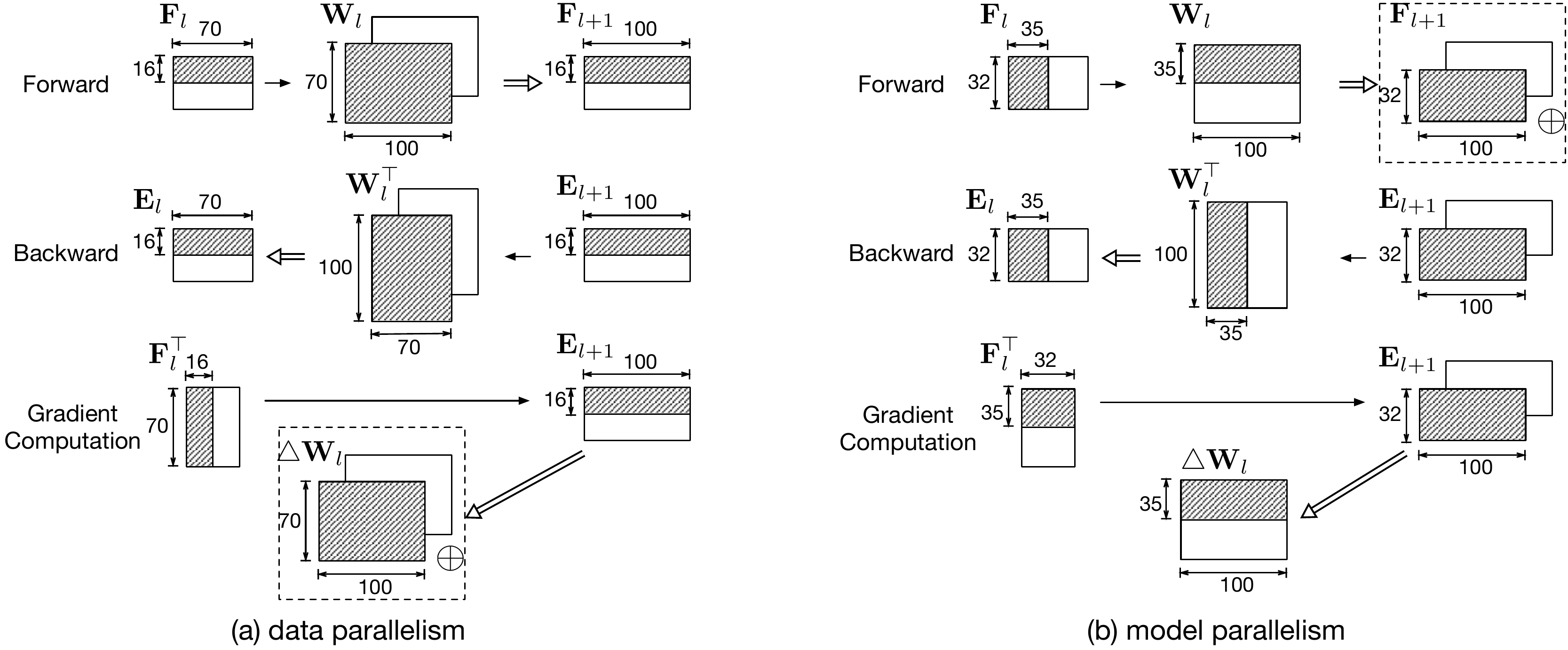}
\vspace{-9pt}
\caption{Forward, Backward and Gradient Computation in (a) data parallelism and (b) model parallelism. In data parallelism, intra-layer communication happens in kernel updating, and in model parallelism intra-layer communication happens in computation for output feature map, both marked by a $\oplus$. }
\label{fig_two_par}
\vspace{-12pt}
\end{figure*}

The goal of \hypar\ is to determine layer-wise parallelism for deep neural network training with an array of DNN accelerators. 
\hypar\ partitions the feature map tensors $\mathbf{F}_{l}$ (input and output) and $\mathbf{F}_{l+1}$, the kernel tensor $\mathbf{W}_{l}$, the gradient tensor $\triangle\mathbf{W}_{l}$, and the error tensors $\mathbf{E}_{l}$ and $\mathbf{E}_{l+1}$ for the DNN accelerators. 
A partition constitutes the choice of parallelism for all weighted layers. 

With mixed data and model parallelism
for each layer, a communication model 
is needed to quantify data movement
amount between accelerators with 
different configuration. 
Applying the insights of communication
in multi-GPU training  system \cite{krizhevsky2014one,tofu,tofu2}, 
we develop a communication model 
between accelerators.
The optimization target is to search a partition that
minimizes the total communication during training a complete deep neural network.
that answers the questions such as: for an 
accelerator and various parallelism settings,
where does the communication come from,
and what is the amount of communication?

Before the technical discussion, 
we clarify some terminologies.
When we use {\em lowercase}  
``data parallelism'', we refer to the case where all
accelerators have a copy of kernel (weight) of 
one specific layer, while feature maps associated with 
that layer are partitioned. When we use {\em lowercase}  
``model parallelism'', we refer to the case where the 
kernels (weights) of one specific layer are
partitioned and 
each accelerator has one partition. We discuss more 
details about data parallelism and model parallelism in 
the following Section \ref{lower_dpmp}. In contrast, we use {\em Uppercase}
``Data Parallelism'' and ``Model Parallelism'' to refer to 
cases where {\em all} layers of a neural network are in 
data parallelism or model parallelism, respectively.

\subsection{Two Types of Parallelism}
\label{lower_dpmp}

We discuss data parallelism and model parallelism using a concrete example. 
Assume we have two accelerators, the batch size is $B=32$. 
Let us consider a fully-connected layer, where 
the number of input and output neurons are 70 and 100, respectively. 
Thus, the feature map $\mathbf{F}_{l}$ has a size of $32\times 70$, the kernel (weight matrix) has a size of $70\times 100$ and $\mathbf{F}_{l+1}$ has a size of $32\times 100$.

\subsubsection{data parallelism}
\label{sec_data_parallelism}

In data parallelism, a batch of data is partitioned into two parts, while the kernels (weight matrix) are duplicated. Each accelerator 
holds one part of the partitioned data and a complete copy of the kernel. 

Figure \ref{fig_two_par} (a) illustrates the shapes of tensors held by the two accelerators. All of the rectangles with shadow lines are held by one accelerator and all of the white rectangles are held by the other. 

In forward, each accelerator performs the computation in Equation \ref{eq_forward}. Because $f(\cdot)$ is an element-wise operation, which only requires local data in the accelerator 
itself but does not require remote data from the other accelerator, 
we focus on the multiplication and represent Equation \ref{eq_forward} as $\mathbf{F}_{l}\to \mathbf{W}_{l} \Rightarrow \mathbf{F}_{l+1}$. For
the one holding the rectangles with shadow lines, 
it performs a multiplication with a size of $[16\times70]\to [70\times100] \Rightarrow [16\times100]$.
Since no remote data are required, 
there is no communication between the two accelerators.

In error backward, the multiplication for each 
accelerator is $\mathbf{E}_{l+1}\to \mathbf{W}^{\top}_{l} \Rightarrow \mathbf{E}_{l}$, and the size of matrices in the multiplication is $[16\times100]\to [100\times70] \Rightarrow [16\times70]$. 
Still, no communication exists between the two accelerators.

However, the remote data access, which leads to 
communication, is required in kernel updating. 
The multiplication is $\mathbf{F}^{\top}_{l}\to\mathbf{E}_{l+1} \Rightarrow \triangle\mathbf{W}_{l}$, and the 
size of matrices in the multiplication in gradient computation is $[70\times16]\to [16\times100] \Rightarrow [70\times100]$. 
Different accelerators compute the gradient with different half of $\mathbf{E}_{l+1}$ and different half of $\mathbf{F}^{\top}_{l}$. 
Therefore, elements in the computed gradient matrix ($[70\times100]$) are 
just partial sums, and the actual gradient is the summation of the two partial 
sums from the two accelerators. 
To update the weights, each accelerator requires remote accesses to the gradient partial sum in the other accelerator, and adds with the local gradient partial sum. We use a $\oplus$ to indicate a remote accesses (and the addition of the partial sum), which incurs communication.

\vspace{-5pt}
\subsubsection{model parallelism}
In model parallelism, the kernel is partitioned, and feature maps are partitioned accordingly.
In forward, each accelerator 
performs computation for the matrices 
$\mathbf{F}_{l}\to \mathbf{W}_{l} \Rightarrow \mathbf{F}_{l+1}$ 
with sizes of $[32\times35]\to [35\times100] \Rightarrow [32\times100]$. 
This scenario is similar to gradient computation in data parallelism.
To get the result for $\mathbf{F}_{l+1}$, remote accesses to the partial sum feature maps from the other accelerator is required. 
In error backward, the multiplication for each 
accelerator is $\mathbf{E}_{l+1}\to \mathbf{W}^{\top}_{l} \Rightarrow \mathbf{E}_{l}$, and the size of matrices in the multiplication is $[32\times100]\to [100\times35] \Rightarrow [32\times35]$. 
No communication exists between the two accelerators.
To compute gradient, the multiplication is $\mathbf{F}^{\top}_{l}\to\mathbf{E}_{l+1} \Rightarrow \triangle\mathbf{W}_{l}$, and the 
size of matrices in the multiplication in gradient computation is $[35\times32]\to [32\times100] \Rightarrow [35\times100]$. 

In model parallelism, the gradient computed by one accelerator is the exact gradient 
needed for updating the kernel held by itself.
Therefore, no communication is required.
However, the communication between two accelerators happens in the computation for output feature maps in forward, which is marked by a $\oplus$ in Figure \ref{fig_two_par} (b).

\subsection{Intra-Layer Communication}
\label{comm_intra}

As shown in Figure \ref{fig_two_par}, each layer performs
three multiplications for forward, error backward, and gradient computation. In each multiplication, three tensors are involved. 
Thus, in total nine tensors need to be considered. 
Notice that $\mathbf{F}_{l}$ and $\mathbf{F}^{\top}_{l}$ are the same,
which is also true for $\mathbf{W}_{l}$ and $\mathbf{W}^{\top}_{l}$, $\mathbf{E}_{l+1}$ and $\mathbf{E}^{\top}_{l+1}$.

For one layer, we call the two tensors on the left hand side of kernel or gradient tensors as \emph{L} tensors; and the two tensors on the right hand side of kernel or gradient tensors as \emph{R} tensors. As shown in Figure \ref{fig_two_par}, $\mathbf{F}_{l}$ ($\mathbf{F}^{\top}_{l}$) and $\mathbf{E}_{l}$ are \emph{L} tensors, and $\mathbf{F}_{l+1}$ and $\mathbf{E}_{l+1}$ are \emph{R} tensors.

Following the idea of intra layer and inter layer from
\cite{song2017pipelayer}, we decouple the communication into two parts:
{\em 1)} \emph{intra-layer communication} by kernel updates within a layer, marked by a $\oplus$ within Figure \ref{fig_two_par}; and 
{\em 2)} \emph{inter-layer communication} by 
conversions of \emph{L} and \emph{R} tensors of feature maps and errors between layers.

In data parallelism, we can see that communication for kernel updating happens when one accelerator remotely accesses $\triangle\mathbf{W}_{l}$ in the other to perform partial sum $\oplus$. We use $\mathbb{A}(\triangle\mathbf{W}_{l})$ to denote the amount of data in $\triangle\mathbf{W}_{l}$. 
Therefore, the intra-layer communication
in data parallelism is $\mathbb{A}(\triangle\mathbf{W}_{l})$, while the 
intra-layer communication in model parallelism is 0. 
In model parallelism, partial sum is to be performed to get $\triangle\mathbf{F}_{l+1}$, so the intra-layer communication is $\mathbb{A}(\mathbf{F}_{l+1})$.  
The intra-layer communication for data parallelism (dp) and 
model parallelism (mp) are summarized in Table \ref{table_in_layer}.

\begin{table}[tb]
\vspace{0pt}
\caption{Intra-layer communication amount in data parallelism and model parallelism.}
\vspace{3pt}
\centering 
\begin{tabular}{|c|c|}
\hline 
data parallelism &model parallelism \\ \hline 
$\mathbb{A}(\triangle\mathbf{W}_{l})$ &
$\mathbb{A}(\mathbf{F}_{l+1})$
 \\ \hline
\end{tabular}
\label{table_in_layer}
\vspace{-12pt}
\end{table}

\subsection{Inter-Layer Communication}
\label{comm_inter}

\begin{figure}[b]
\centering
\vspace{-12pt}
\includegraphics[width=0.9\columnwidth]{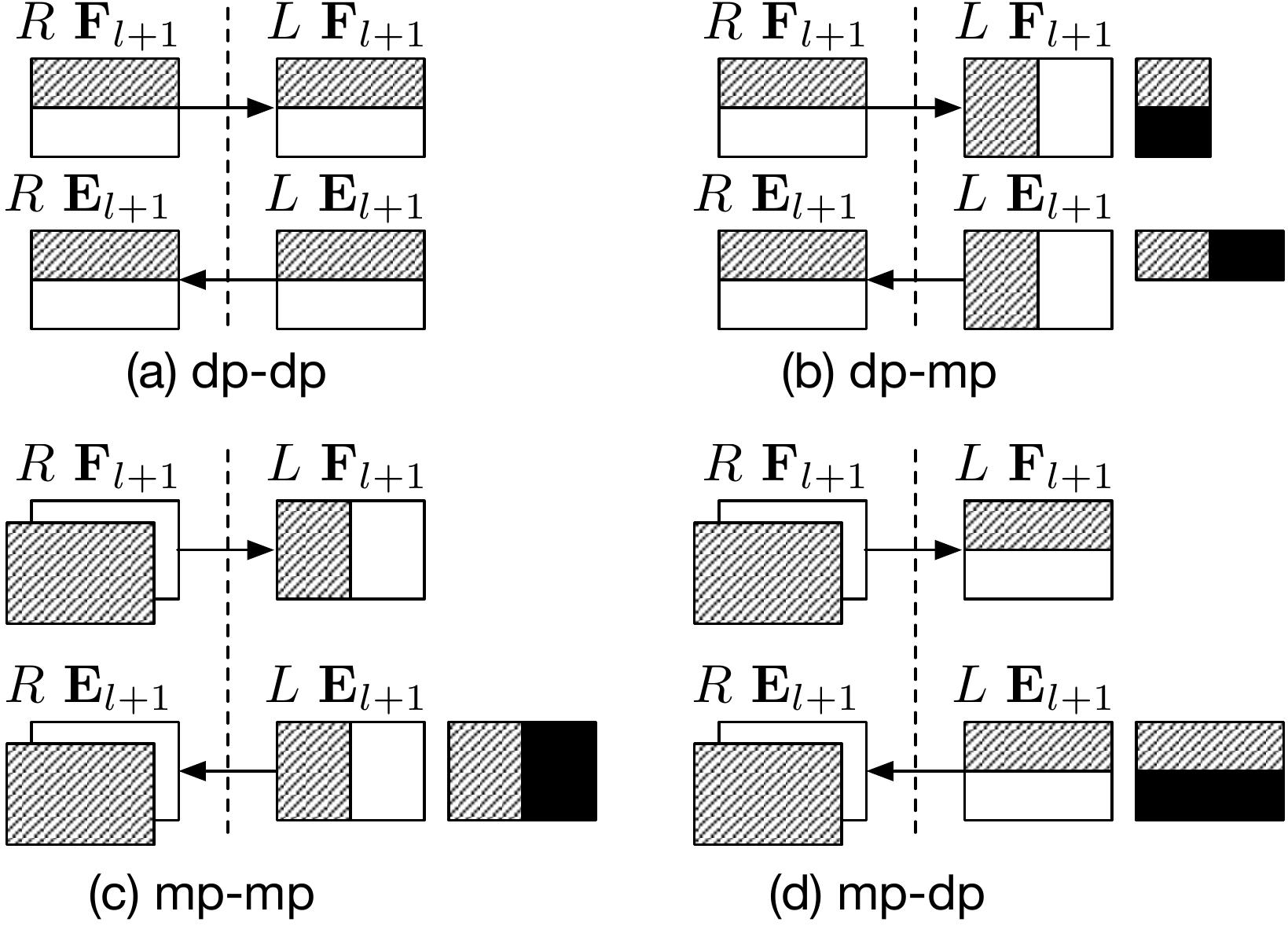}
\vspace{-9pt}
\caption{Inter-layer communication for (a) dp-dp, (b) dp-mp, (c) mp-mp and (d) mp-dp.}
\label{fig_inter_layer_com}
\vspace{0pt}
\end{figure}

Next, we calculate the inter-layer communication due to accessing 
feature maps and errors. Essentially, we calculate 
the communication for conversion of \emph{L} and \emph{R} tensors, as shown in Figure \ref{fig_inter_layer_com}.
Since the parallelism for each layer is either data parallelism or model parallelism, to calculate the communication for each layer, we should consider four cases: dp-dp, dp-mp, mp-mp and mp-dp.

\vspace{3pt}
\noindent
\textbf{dp-dp} In Figure \ref{fig_inter_layer_com} (a), the \emph{R} tensors $\mathbf{F}_{l+1}$ and $\mathbf{E}_{l+1}$ belong to a previous layer, Layer $l$, and \emph{L} tensors belong to Layer $l+1$. For the accelerator which holds the dashed-line tensors, no remote access to the accelerator is necessary because the \emph{R} and \emph{L} tensors have the same shape. Thus, the inter-layer communication in dp-dp is 0.

\vspace{3pt}
\noindent  
\textbf{dp-mp} 
The \emph{R} and \emph{L} tensors in Figure \ref{fig_inter_layer_com} (b) have different shapes, which causes communication between two accelerators. For the accelerator which holds the dashed-line tensors, it needs the \emph{L} tensor $\mathbf{F}_{l+1}$, but the \emph{R} tensor $\mathbf{F}_{l+1}$ held by this accelerator (from the computation it performs on Layer $l$) has a different shape compared to the \emph{L} tensor $\mathbf{F}_{l+1}$. 
Thus, this accelerator needs to remotely read part of the white \emph{R} tensor $\mathbf{F}_{l+1}$ from the other accelerator. The shape of the communication is the overlaps of the white \emph{R} tensor $\mathbf{F}_{l+1}$ and the dashed-line \emph{L} tensor $\mathbf{F}_{l+1}$, i.e., the black tensor. The black tensor is 1/4 of the tensor $\mathbf{F}_{l+1}$, so the inter-layer communication 
between \emph{R} $\mathbf{F}_{l+1}$ and \emph{L} $\mathbf{F}_{l+1}$ is $0.25\mathbb{A}(\mathbf{F}_{l+1})$. Similarly, we can calculate
and obtain that the inter-layer
communication 
between \emph{R} $\mathbf{E}_{l+1}$ and \emph{L} $\mathbf{E}_{l+1}$ is $0.25\mathbb{A}(\mathbf{E}_{l+1})$.
Therefore, the inter-layer communication
in dp-mp is $0.25\mathbb{A}(\mathbf{F}_{l+1}) + 0.25\mathbb{A}(\mathbf{E}_{l+1})$.

\vspace{3pt}
\noindent 
\textbf{mp-mp}
Because the dashed-line \emph{R} $\mathbf{F}_{l+1}$ already contains 
the dashed-line \emph{L} $\mathbf{F}_{l+1}$, the communication for $\mathbf{F}_{l+1}$ is 0. But for $\mathbf{E}_{l+1}$, the dashed-line \emph{R} $\mathbf{E}_{l+1}$ requires remote access to obtain the black part, the communication is $0.5\mathbb{A}(\mathbf{E}_{l+1})$.

\vspace{3pt}
\noindent 
\textbf{mp-dp}
Similar to mp-mp, we can calculate that the inter-layer communication for mp-dp is $0.5\mathbb{A}(\mathbf{E}_{l+1})$.

The results based on the above inter-layer communication calculation are 
summarized in Table \ref{table_inter_layer}.

\begin{table}[tb]
\centering 
\caption{Inter-layer communication amount for the transition of dp-dp, dp-mp, mp-mp and mp-dp.}
\vspace{3pt}
\begin{tabular}{|c|c|}
\hline 
dp-dp &0 \\ \hline 
dp-mp & $0.25\mathbb{A}(\mathbf{F}_{l+1}) + 0.25\mathbb{A}(\mathbf{E}_{l+1})$ \\ \hline 
mp-mp & $0.5\mathbb{A}(\mathbf{E}_{l+1})$ \\ \hline
mp-dp & $0.5\mathbb{A}(\mathbf{E}_{l+1})$
 \\ \hline
\end{tabular}
\label{table_inter_layer}
\vspace{-12pt}
\end{table}

From Table \ref{table_in_layer} and Table \ref{table_inter_layer},
we can see that, for DNN inference, the best option is Data Parallelism, i.e., data parallelism for every layers.
It is because 
the intra-layer communication is zero since no gradient computation is necessary in inference, and the inter-layer communication of dp-dp is also zero. 
However, the assumption is different for DNN training, 
and parallelism becomes a critical concern.

\subsection{Parallelism in Training}
\label{comm_discuss}

We use the example in Figure \ref{fig_two_par} to compare the communication of data parallelism and model parallelism in our communication model.
In Figure \ref{fig_two_par}, for a fully-connected layer, 
we see that the communication amount between the two accelerators in data parallelism is 56KB($=2\times70\times100\times4$B) 
assuming a precision of 32-bit floating point and a batch size of 32. 
The communication amount in model parallelism is 25.6KB($=2\times32\times100\times4$B). 
In this case,  model parallelism is better than data parallelism. 
However, if we consider computations in a convolutional layer, where the $\mathbf{F}_{l}$ has a size of $[12\times12\times20]$, $\mathbf{W}_{l}$ has a size of $[5\times5\times20]\times50$ and $\mathbf{F}_{l+1}$ has a size of $[8\times8\times50]$, the communication amount in data parallelism is 200KB($=2\times5\times5\times20\times50\times4$B), while communication
amount in model parallelism is 819KB($=2\times32\times8\times8\times50\times4$B). 
In this different scenario, data parallelism is better than model parallelism.
From the simple example, we see that a 
default approach that choose either data parallelism or model parallelism 
for all layers in a network would not get the highest performance because both convolutional and fully-connected layers exist in most widely used deep neural networks.
That also explains why fully-connected layers are configured with model parallelism and convolutional layers are configured with data parallelism in the empirical ``one weird trick'' \cite{krizhevsky2014one}, which outperforms default Data Parallelism and Model Parallelism.

However, the empirical configuration in \cite{krizhevsky2014one} will {\em not} always work. With our
general communication model, we can see the trick \cite{krizhevsky2014one} only considered intra-layer communication,
but did not 
consider the other communication source, inter-layer communication.  We will show that \hypar\ performs better than the trick in Section \ref{sec_hypar_to_the_trick}.

Realizing this fact, one may naturally choose a more comprehensive approach:
choosing a parallelism for each layer and enumerating all possibilities 
to determine the best choice. 
Unfortunately, it is not feasible, because the time complexity for such 
enumeration is $O(2^N)$ for a neural network with $N$ weighted layers.
To come up with a practical solution, 
we apply the dynamic programming-based
search and recursive partitioning method proposed in \cite{tofu,tofu2}  which partitions
on tensors and feature maps, respectively.
In our case, we partition on
a set of coalesced tensors, i.e., either data or model parallelism.


%% file: sec_partition.tex
In this section, we discuss \hypar, which partitions the feature map tensors $\mathbf{F}_{l}$ (input and output) and $\mathbf{F}_{l+1}$, the kernel tensor $\mathbf{W}_{l}$, the gradient tensor $\triangle\mathbf{W}_{l}$, and the error tensors $\mathbf{E}_{l}$ and $\mathbf{E}_{l+1}$ for the DNN accelerators. 
The parallelism for one layer actually determines the tensor partitioning for two accelerators, as shown in Figure \ref{fig_two_par}.
A partition constitutes the choice of parallelism for each layer in a deep neural network. 
The optimization target is to search a partition that
minimizes the total communication during training a complete deep neural network.
Rather than $O(2^N)$ brute-force search, \hypar\ is practical: the time complexity for the partition 
search in \hypar\ is {\em linear}, i.e. $O(N)$ for a neural network with $N$ weighted layers.

\subsection{Partition Between Two Accelerators}

From Section \ref{comm_inter}, we can see that:
{\em 1)} each layer is configured with either data parallelism or model parallelism;
{\em 2)} the calculation for inter-layer communication only depends on two adjacent layers;
{\em 3)} the intra-layer communication only depends on the parallelism of that layer, but does not depend on any other layer. Thus, to minimize the total
amount of communication, we can use a {\em layer-wise dynamic programming} method to search for the partitions for each layer. The time complexity of this method is $O(N)$ for a neural network with $N$ weighted layers.

\begin{algorithm}[htb] 
\small

\caption{ Partition Between Two Accelerators.} 
\label{alg_partition_two} 
\begin{algorithmic}[1] 
\REQUIRE ~~\\ 
1. Batch size, $B$,\\
2. The number of weighted layers in a DNN model, $L$,\\
3. A list of hyper parameters (layer type: conv or fc, kernel sizes, parameter for pooling, activation function), $HP[l], l=0,..,L$-$1$.
\ENSURE ~~\\ 
1. Total communication, $com$,\\
2. A list of parallelism for each weighted layer, $P[l], l=0,..,L$-$1$.\\
\vspace{9pt}
\STATE Generate tensor shapes for $\mathbf{F}_{l}$, $\mathbf{W}_{l}$, $\triangle\mathbf{W}_{l}$, $\mathbf{E}_{l}$ for each layer.
\STATE Initiate $com\_dp[0]$ = $0$, $com\_mp[0]$ = $0$, $P\_dp =$[~], \\ $P\_mp =$[~].
\FOR{($l=0$; $l<L$; $l$++)}  
\STATE Compute $intra\_dp$, $intra\_mp$ using Table \ref{table_in_layer} and \\
$inter\_dp\_dp$, $inter\_mp\_dp$, $inter\_dp\_mp$,\\ $inter\_mp\_mp$ using Table \ref{table_inter_layer}.
\STATE $com\_dp[l]$ = min($com\_dp[l$-$1]$+$inter\_dp\_dp$, \\ 
~~~~~~~~~~~~~$com\_mp[l$-$1]$+$inter\_mp\_dp$) + $intra\_dp$.
\STATE Update $P\_dp$.
\STATE $com\_mp[l]$ = min($com\_dp[l$-$1]$+$inter\_dp\_mp$,\\
~~~~~~~~~~~~~$com\_mp[l$-$1]$+$inter\_mp\_mp$)+$intra\_mp$.
\STATE Update $P\_mp$.
\ENDFOR

\RETURN min($com\_dp[L$-$1]$, $com\_mp[L$-$1]$) \\
and the corresponding parallelism list ($P\_dp$ or $P\_mp$).
\end{algorithmic}
\end{algorithm}

The idea is to use dynamic programming, for each layer, to compute the intra-layer communication of dp and mp and inter-layer communication of dp-dp, mp-dp, dp-mp, mp-mp using the results in 
Table \ref{table_in_layer} and Table \ref{table_inter_layer}, and 
then calculate the minimum {\em accumulated} communication for data parallelism or model parallelism in this layer. 


The pseudocode of the partition algorithm between two accelerators is given in Algorithm~\ref{alg_partition_two}. 
The inputs of our partition algorithm are identical to the parameters that are needed for a normal mini batch training process. As shown in the input section of Algorithm~\ref{alg_partition_two}, the inputs include the batch size ($B$), the number of model layers ($L$), and the necessary hyper parameters (layer type: conv or fc, kernel sizes, parameter for pooling, activation function: $HP[l], l=0,..,L-1$).
The outputs of our partition algorithm are composed of the minimal total communication between two accelerators and a list parallelism methods we should chose to realize such minimal communication for each layer in the model.

\subsection{Hierarchical Partition}
So far, we assume only two accelerators. 
To expand to partition for an array of accelerators, 
we use a hierarchical approach.

Although Algorithm \ref{alg_partition_two} performs partitions 
based on two accelerators, we can view the two accelerators as two groups of accelerators. Then, the Algorithm \ref{alg_partition_two} can be used to partition between two groups. 
Based on this insight, we have a hierarchical partition algorithm.

\begin{figure}[tb]
\vspace{-0pt}
\centering
\includegraphics[width=0.75\columnwidth]{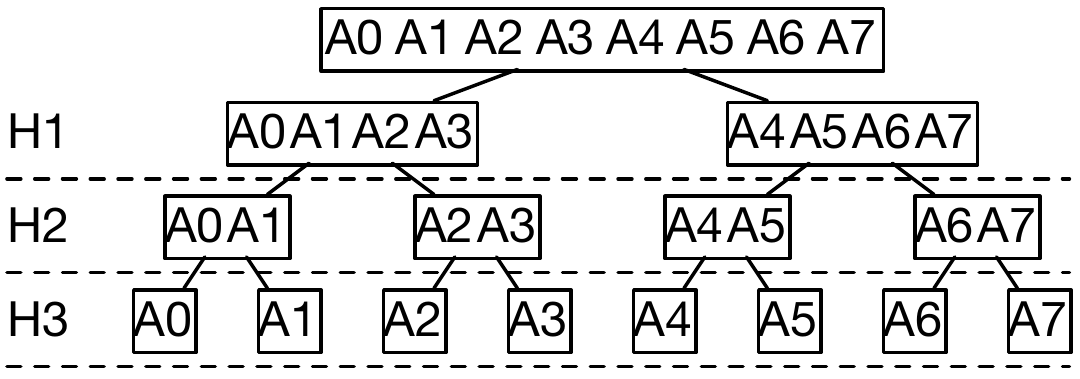}
\vspace{-9pt}
\caption{Hierarchical partition for three levels.}
\label{fig_H_par}
\vspace{-12pt}
\end{figure}


Figure \ref{fig_H_par} illustrates an example to partition eight accelerators with three levels. 
Originally, there are $8$ accelerators waiting to be assigned with the training workloads.
As shown in Figure~\ref{fig_H_par}, at hierarchy level H1, we first view A0 to A3 as a group of accelerators and A4 to A7 as the other group. By utilizing Algorithm \ref{alg_partition_two}, the workloads can be assigned to each of these two groups.
After that, the groups of A0 to A3 and A4 to A7 are further divided in to $4$ groups at hierarchy level H2. Last, as illustrated in Figure~\ref{fig_H_par}, each accelerator is assigned its own workloads at hierarchy level H3 with the same partition method.
Such binary tree structure of our hierarchical partition method enables the partition of $8$ accelerators in logarithmic times of iteration, which equals $3$ ($log_{2}{8}$).
With Figure \ref{fig_H_par}, at each hierarchy level, we have a parallelism list for every weighted layers for layers to be partitioned into the two subarrays. In total three parallelism lists are generated by the partition algorithm and we can use the lists to determine the parallelism setting for each accelerator, and the tensors they hold.


\begin{algorithm}[htb] 
\small
\caption{ Hierarchical Partition.} 
\label{alg_partition_tH} 
\begin{algorithmic}[1] 
\REQUIRE ~~\\ 
1. The number of hierarchy levels, $H$,\\
2. Batch size, $B$,\\
3. The number of weighted layers in a DNN model, $L$,\\
4. A list of hyper parameters (layer type: conv or fc, kernel sizes, parameter for pooling, activation function), $HP[l], l=0,..,L$-$1$.
\ENSURE ~~\\ 
1. Total communication, $com$,\\
2. A list of parallelism for each weighted layer at each hierarchy level, $P[h][l], h=0,...,H$-$1, l=0,..,L$-$1$.\\
\vspace{9pt}
\IF{($H$ == 0)}
\RETURN (0, [~~]).
\ELSE
\STATE ($com\_h, P\_h$) = PartitionBetweenTwoAccelerators().
\STATE ($com\_n, P\_n$) = HierarchicalPartition($H$-$1$).
\STATE Update $P$.
\STATE $com=com\_h + 2*com\_n$.
\RETURN ($com$, $P$).
\ENDIF
\end{algorithmic}
\end{algorithm}

In the hierarchical partition algorithm, for one specific hierarchy, we first partition an array of the accelerators into two subarrays by Algorithm \ref{alg_partition_two}, and then recursively apply the hierarchical partition algorithm to the subarray until there is only one accelerator in one subarray. The hierarchical partition can be summarized as Algorithm \ref{alg_partition_tH}.


The inputs of Algorithm \ref{alg_partition_tH} are similar to those of Algorithm \ref{alg_partition_two} except that the hierarchical partition algorithm also needs the hierarchy levels($H$) as the input to denote how many times we divide the training workload into two groups. For example, if hierarchy levels $H$ equals 4, the total number of accelerators in the array are $2^{H}$. The outputs of hierarchical partition algorithm include the total communication through all the hierarchy levels ($com$) and a list containing all the partition strategies for each layer of all the hierarchy levels in the accelerator array.

Our hierarchical partition algorithm is a recursive function. In each recursion, Algorithm \ref{alg_partition_tH} first calculates the minimal communication ($com\_h$) for current hierarchy level using \ref{alg_partition_two}. Then, it calls itself with the input hierarchy levels($h$) changed to hierarchy levels($h-1$) and obtains the total minimal communication ($com\_n$) for the lower hierarchy levels. At last it returns the total communication $com$ by adding current communication $com\_h$ and $2\times com\_n$.




{\bf Discussion}.
While we applied the general method in \cite{tofu} in our parallelism search, 
the detail of our algorithm is
different in one important way.
When performing the matrix multiplication (two input tensors, one output tensor), \cite{tofu} considers three partitions for each operator and two for each tensor in forward and backward, leading to several hundred combinations {\em for one layer}. In HyPar, our assumption is that the partition of tensors in the three phases (i.e., forward, backward and gradient computing) are coalesced as we only considered 2 cases (model/data parallelism). Thus, our searching space is much smaller.
The downside of shrinking the search space is that the resulting parallel configuration may be 
worse in performance than that found in \cite{tofu} which explores a larger search space.

%% file: sec_implementation.tex


\begin{figure}[t]
\centering
\vspace{-0pt}
\includegraphics[width=0.8\columnwidth]{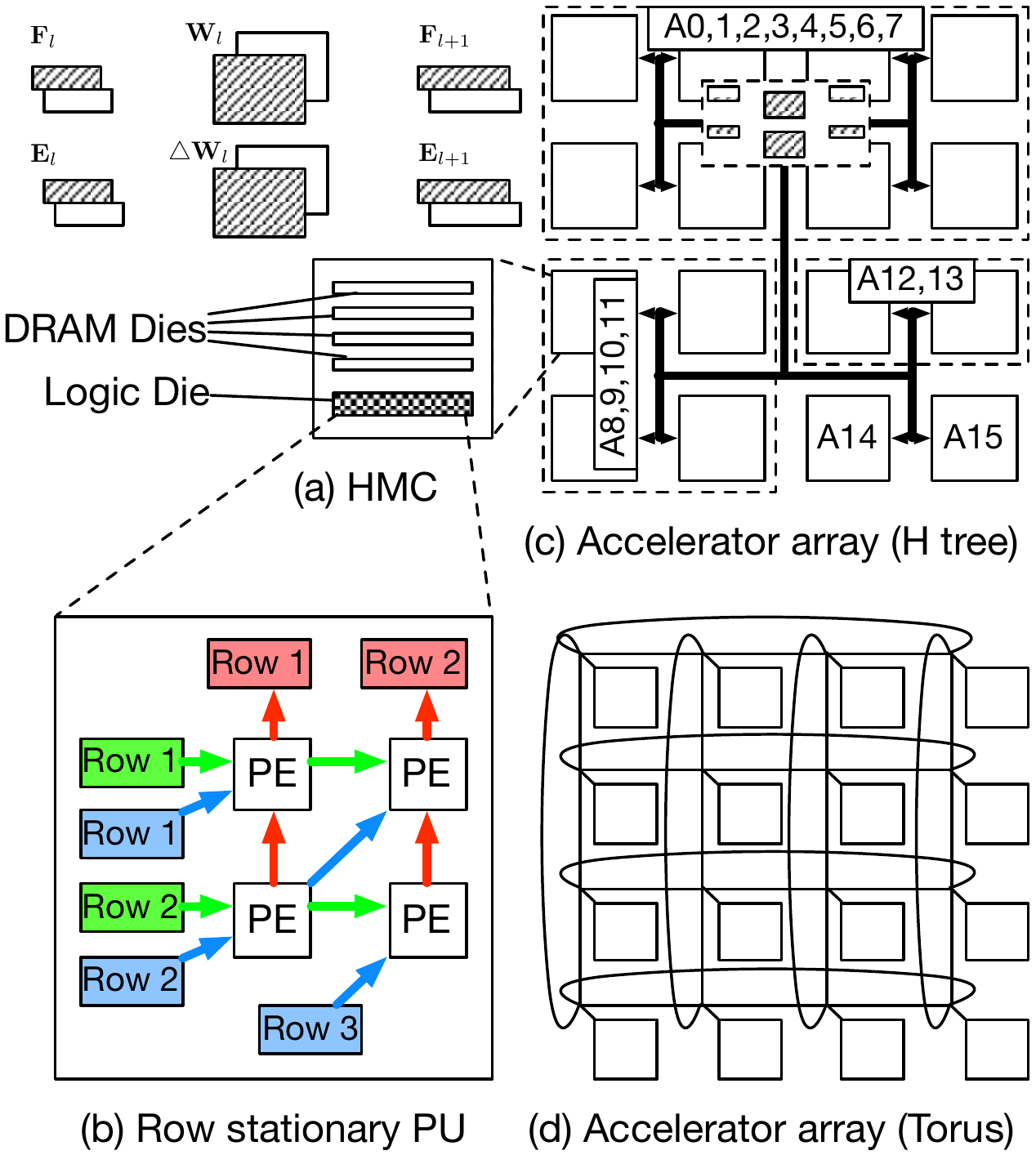}
\vspace{-9pt}
\caption{Overall view of (a) an HMC-based accelerator, (b) a row stationary processing unit, (c) an array of sixteen accelerators in H tree, and (d) the accelerator array in torus.}
\label{fig_hypar_arch}
\vspace{-12pt}
\end{figure}

\begin{figure*}[tb]
\vspace{-0pt}
\centering
\includegraphics[width=1.95\columnwidth]{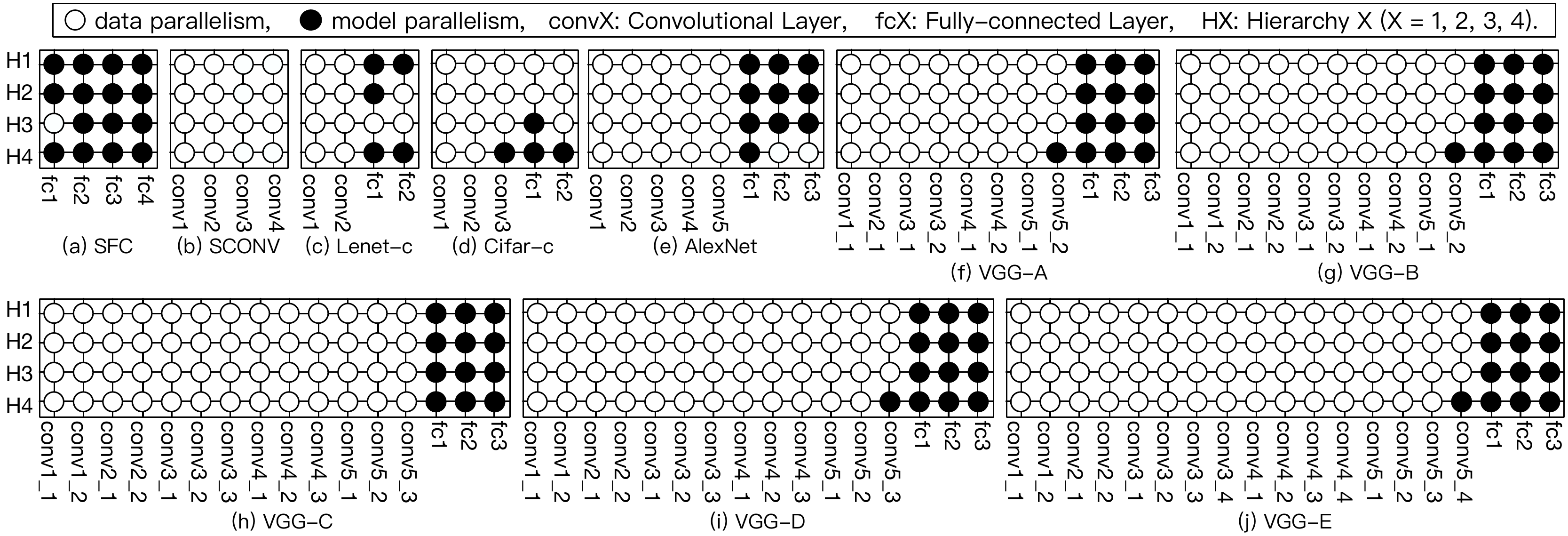}
\vspace{-9pt}
\caption{Optimized parallelism for weighted layers in four hierarchy levels of ten networks in \hypar. }
\label{fig_opt_par}
\vspace{-12pt}
\end{figure*}

This section presents the \hypar\ architecture 
composed of an accelerator array, where the 
parallelism setting is determined by \hypar.
The individual accelerator is based on Hybrid Memory Cube (HMC)~\cite{hmcspec}, as shown in Figure \ref{fig_hypar_arch} (a). 
An HMC consists of stacked DRAM dies and logic die, and they are connected by through silicon vias (TSVs),
Processing units (PUs) can be integrated on the logic die.
HMC provides high memory bandwidth (320 GB/s) \cite{hmcspec}, which is suitable for the in-memory processing for DNNs.
Since DNN accelerators incur heavy memory accessing and intensive computation operations, recent works \cite{kim2016neurocube, gao2017tetris} 
demonstrated that HMC-based neural network accelerator could 
drastically reduce data movements.

For the PUs, as shown in Figure \ref{fig_hypar_arch} (b), 
we implement a row stationary design as \cite{chen2017eyeriss}. 
In such design, weight rows (green) are shared by processing engines horizontally,
feature map rows (blue) are shared by processing engines diagonally, and
partial sum rows (red) are accumulated vertically. The row stationary design is suitable for convolution computations.

Figure \ref{fig_hypar_arch} shows the overall \hypar\ architecture composed of
a 2-D array of sixteen accelerators.
Thus, the number of hierarchy levels for the accelerator array is four. 
To support efficient hierarchical communication,
the accelerators in the array are connected with 
certain network topology. 
Notice that \hypar\ algorithm hierarchically partitions the accelerators. 
In each level, the two subarrays
of accelerators (or two accelerators in the last level) 
communicate between each other. The communication relationship is represented by the two edges from their ancestor subarray. For example, as shown in Figure \ref{fig_H_par}, in level H2, the ancestor subarray \{A0, A1, A2, A3\} has edges connected to the child subarray \{A0, A1\} and \{A2, A3\}. That means subarray \{A0, A1\} and \{A2, A3\} communicate with each other. For subarray \{A0, A1\} to communicate with subarray \{A4, A5\}, the communication ``backtracks'' to level H1. That is natural because in the hierarchical partition Algorithm \ref{alg_partition_tH}, subarray \{A0, A1\} and \{A4, A5\} have no direct communication in level H2, but they may communicate in level H1. 

We consider two network topologies. 
Figure \ref{fig_hypar_arch} (c) shows the
H tree topology, which can match the communication patterns. As shown in Figure \ref{fig_hypar_arch} (b), the tensors with shadow lines are assigned to subarray \{A0-7\}, while the white tensors are assigned to subarray \{A8-15\}, and the tensors in each subarray are recursively assigned to sub-subarrays according to the hierarchical partition.

Figure \ref{fig_hypar_arch} (d) shows the torus topology for the accelerator array. 
While torus is a common topology, it performs worse 
than the H tree. It is because the tensor partition pattern 
generated by Algorithm \ref{alg_partition_tH} 
does not match torus topology as well as the H tree. 

%% file: sec_eval.tex
\subsection{Evaluation Setup}

In the evaluation, we use three datasets, including small, medium and large size:
MNIST \cite{lecun1998mnist}, CIFAR-10 \cite{krizhevsky2014cifar} and ImageNet \cite{deng2009imagenet}. We use ten deep neural network models in the evaluation: {\tt SFC}, {\tt SCONV}, {\tt Lenet-c}, {\tt Cifar-c}, {\tt AlexNet}, {\tt VGG-A}, {\tt VGG-B}, {\tt VGG-C}, {\tt VGG-D} and {\tt VGG-E}. 
{\tt SFC} is a fully-connected network (no convolutional layers), and {\tt SCONV} is a convolutional network without any fully-connected layers. The hyper parameters of {\tt SFC} and {\tt SCONV} are shown in Table \ref{table_two_network_parameter}. Notice that {\tt SFC} and {\tt SCONV} are valid networks, which have an accuracy of 98.28\% and 98.71\%, respectively. We build the two ``strange'' networks as extreme cases to demonstrate 
the effects of data and model parallelism.
{\tt Lenet-c} is a convolutional neural network for MNIST, {\tt Cifar-c} is for Cifar-10. {\tt AlexNet} and {\tt VGG}s are for ImageNet, and the model hyper parameters can be found in \cite{alexnet2012} and \cite{simonyan2014very} respectively.

\begin{table}[tb]
\centering 
\vspace{-6pt}
\caption{Hyper parameters for {\tt SFC} and {\tt SCONV}.}
\vspace{3pt}
\begin{tabular}{||c|l||}
\hline 
{\tt SFC} &784-8192-8192-8192-10 \\ \hline 
\multirow{2}{*}{{\tt SCONV}} &20@5$\times$5, 50@5$\times$5(2$\times$2 max pool), \\
 &50@5$\times$5, 10@5$\times$5(2$\times$2 max pool) \\\hline 
\end{tabular}
\label{table_two_network_parameter}
\vspace{-12pt}
\end{table}

For the accelerator array, we employ sixteen accelerators
(as shown in Figure~\ref{fig_hypar_arch}). The number of partition hierarchy levels is four. Each accelerator is based on an HMC cube. The batch size is 256. 
We use an event-driven simulation. Within an HMC vault (i.e., an Eyeriss accelerator and its local memory), we modeled the computation cost and the memory access between vaults, we also considered the tensor communication. 
For the HMC, the DRAM bandwidth is 320 GB/s and each HMC has 8 GB memory \cite{hmcspec}. The PUs used in the evaluation have an Eyeriss-like \cite{chen2017eyeriss} row stationary architecture, and each processing unit has 168 (12$\times$14) processing engines, 108 KB on-chip buffer and 84.0 GOPS/s computation density. The accelerator works at 250 MHz and the link bandwidth is 1600 Mb/s (i.e. the total network bandwidth is 25.6 Gb/s).
The energy consumption for a 32-bit float ADD operation is 0.9 pJ, a 32-bit float MULT operation is 3.7 pJ, a 32-bit SRAM accessing is 5.0 pJ and a 32-bit DRAM accessing is 640 pJ \cite{horowitz20141}. We use a precision of 32-bit floating point in the computation.

We compare the default Model Parallelism (where all layers at the four hierarchy levels are assigned to model parallelism), the default Data Parallelism (where all layers at the four hierarchy levels are assigned to data parallelism) and \hypar\ in the evaluation.

\subsection{Overall Results}
\subsubsection{Optimized Parallelism in \hypar}

Figure \ref{fig_opt_par} shows the optimized parallelisms for weighted layers in the ten networks at four hierarchy levels. For most networks, especially large-scale networks, such as {\tt AlexNet} and {\tt VGG}s, in the convolutional layers, the parallelisms are usually data parallelism, and in fully-connected layers, the parallelisms usually are model parallelism. That is consistent to our analysis in Section \ref{comm_intra}, i.e., convolutional layers favor data parallelism while fully connected layers prefer model parallelism to keep the communication as low as possible. The optimization of extreme cases is a slightly different.
For {\tt SFC}, because all layers are fully-connected layer, except {\tt fc1}@H3 is optimized to data parallelism, all other layers at the four hierarchy levels are optimized to model parallelism. For {\tt SCONV}, a network with all convolutional layers, all layers at the four hierarchy levels are optimized to data parallelism. We also see that except {\tt SCONV}, the optimized parallelisms for layers at four hierarchy levels consist of both data parallelism and model parallelism, leading to hybrid parallelism.

\subsubsection{Performance}

\begin{figure}[b]
\centering
\vspace{-9pt}
\includegraphics[width=0.90\columnwidth]{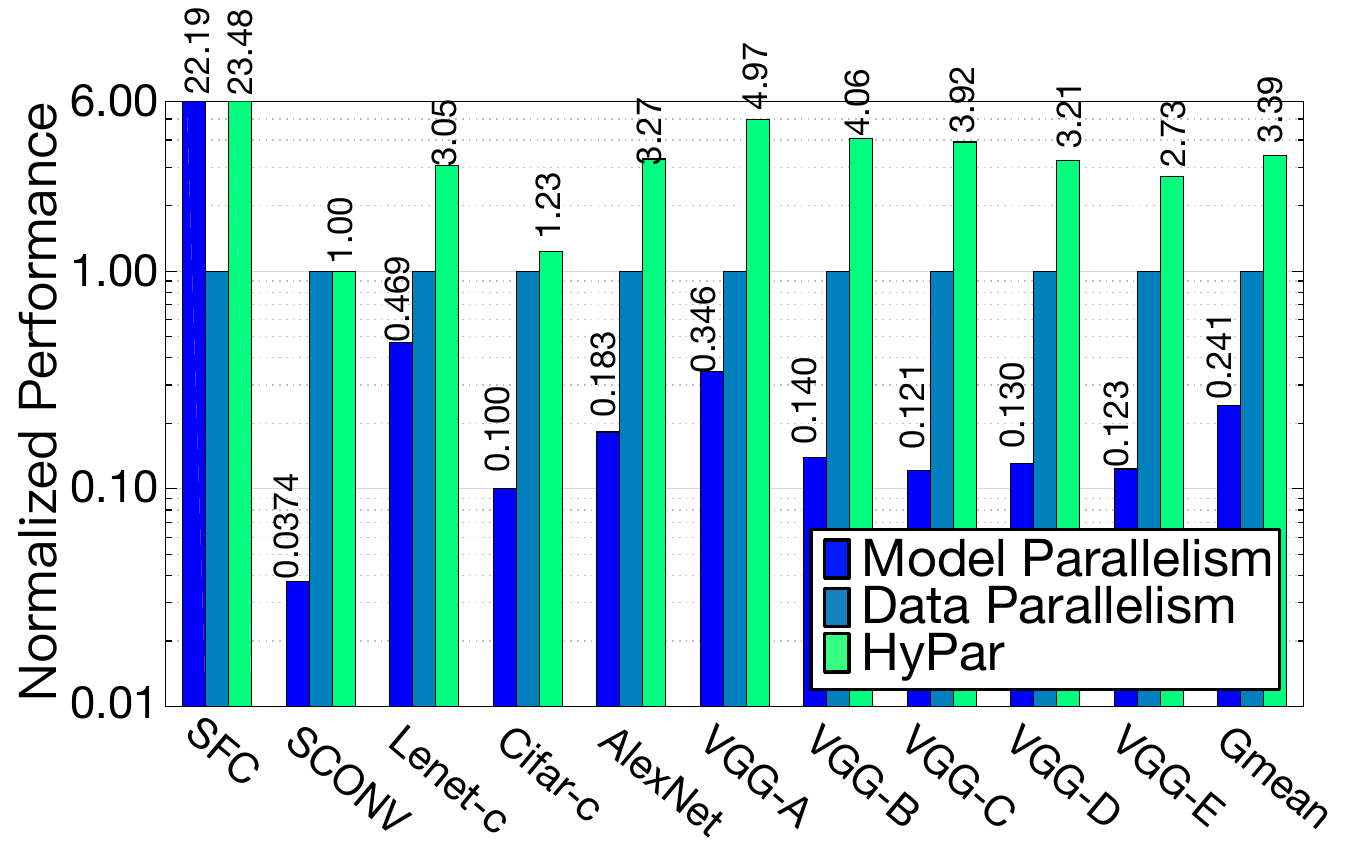}
\vspace{-9pt}
\caption{Performance of Model Parallelism, Data Parallelism and \hypar\ normalized to Data Parallelism.}
\label{fig_eval_perf}
\vspace{-0pt}
\end{figure}

The performance of the default Model Parallelism, the default Data Parallelism and \hypar\ are shown in Figure \ref{fig_eval_perf}. The performance results are normalized to the default Data Parallelism. 

\hypar\ achieves a 3.39$\times$ performance gain compared to Data Parallelism on average. We can also find that the performance of
Model Parallelism is almost always worse than Data Parallelism. 
Thus, among these two, we should mostly prefer Data Parallelism in DNN training. For the extreme case {\tt SFC}, Model Parallelism performs better than Data Parallelism, but \hypar\ still performs slightly better than Model Parallelism. Although fully-connected layers prefer Model Parallelism, as shown in Figure \ref{fig_opt_par} (a), {\tt fc1}@H3 is optimized to Data Parallelism.
Therefore, the optimized parallelisms in \hypar\ are not fully Model Parallelism, and this explains why \hypar\ performs better than Model Parallelism (23.48$\times$ v.s. 22.19$\times$). 
It validates the partitioning algorithm of \hypar. For the other extreme case {\tt SCONV}, \hypar\ performs the same as Data Parallelism. For other eight networks, \hypar\ achieves performance gains ranging from 1.23$\times$ to 4.97$\times$ compared to Data Parallelism.

\subsubsection{Energy Efficiency}

\begin{figure}[tb]
\centering
\vspace{-0pt}
\includegraphics[width=0.95\columnwidth]{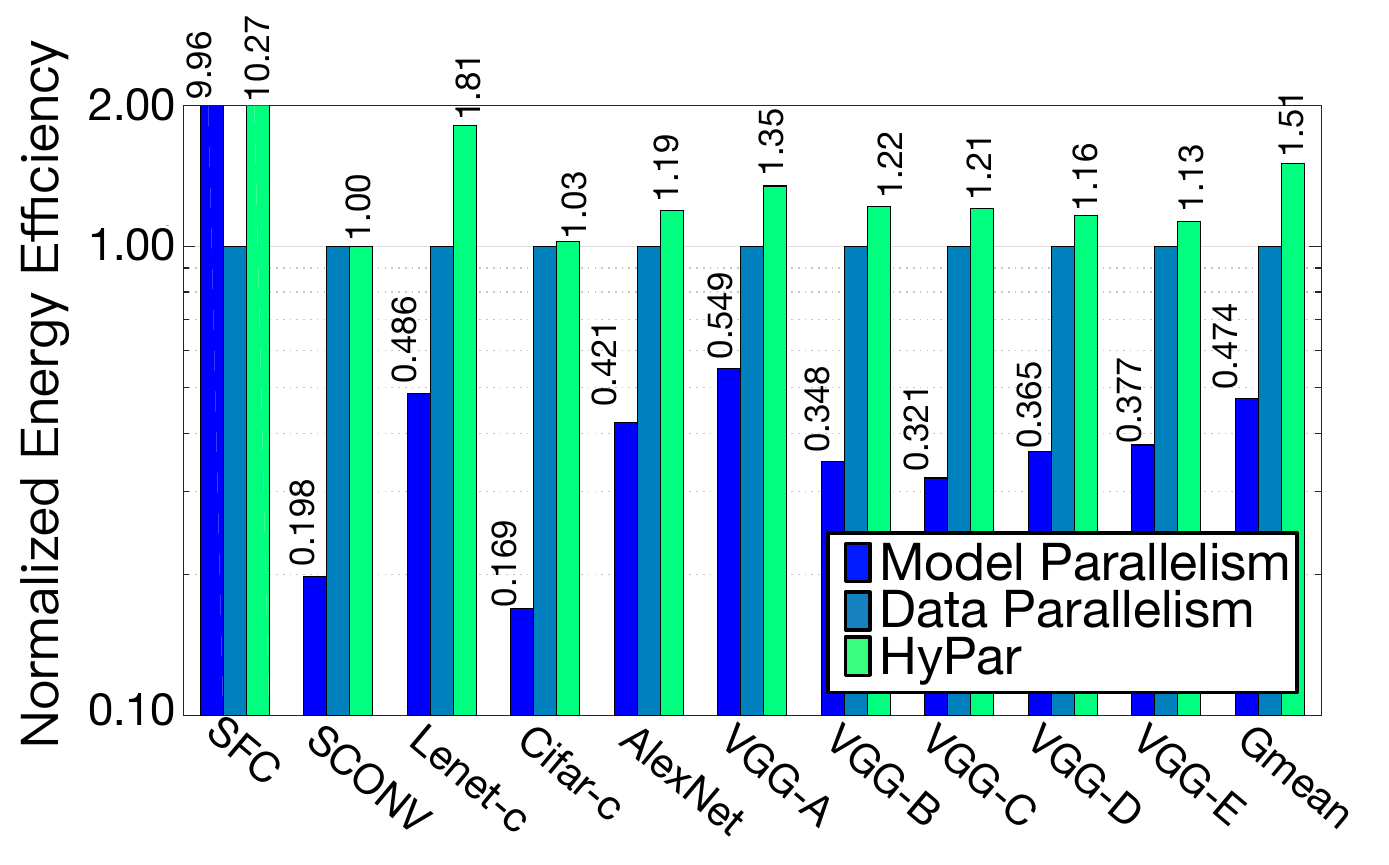}
\vspace{-9pt}
\caption{Energy efficiency of Model Parallelism, Data Parallelism and \hypar\ normalized to Data Parallelism.}
\label{fig_eval_energy}
\vspace{-9pt}
\end{figure}

The energy efficiency of the default Model Parallelism, the default Data Parallelism and \hypar\ are shown in Figure \ref{fig_eval_energy}. The energy efficiency is the the energy saving normalized to the default Data Parallelism. 

\hypar\ achieves a 1.51$\times$ energy efficiency compared to Data Parallelism on average. Again, Model Parallelism is almost always less energy efficiency than Data Parallelism. For the extreme case {\tt SFC}, the energy efficiency of Model Parallelism is higher than that of Data Parallelism, but \hypar\ performs better than Model Parallelism, and \hypar\ has higher energy efficiency (10.27$\times$) than Model Parallelism (9.96$\times$). For the other extreme case {\tt SCONV}, \hypar\ has a 1.00$\times$ energy efficiency, the same as Data Parallelism. For other eight networks, \hypar\ achieves energy efficiencies ranging from 1.03$\times$ to 1.81$\times$ compared to Data Parallelism.

\subsubsection{Total Communication per Step}

\begin{figure}[b]
\centering
\vspace{-9pt}
\includegraphics[width=0.9\columnwidth]{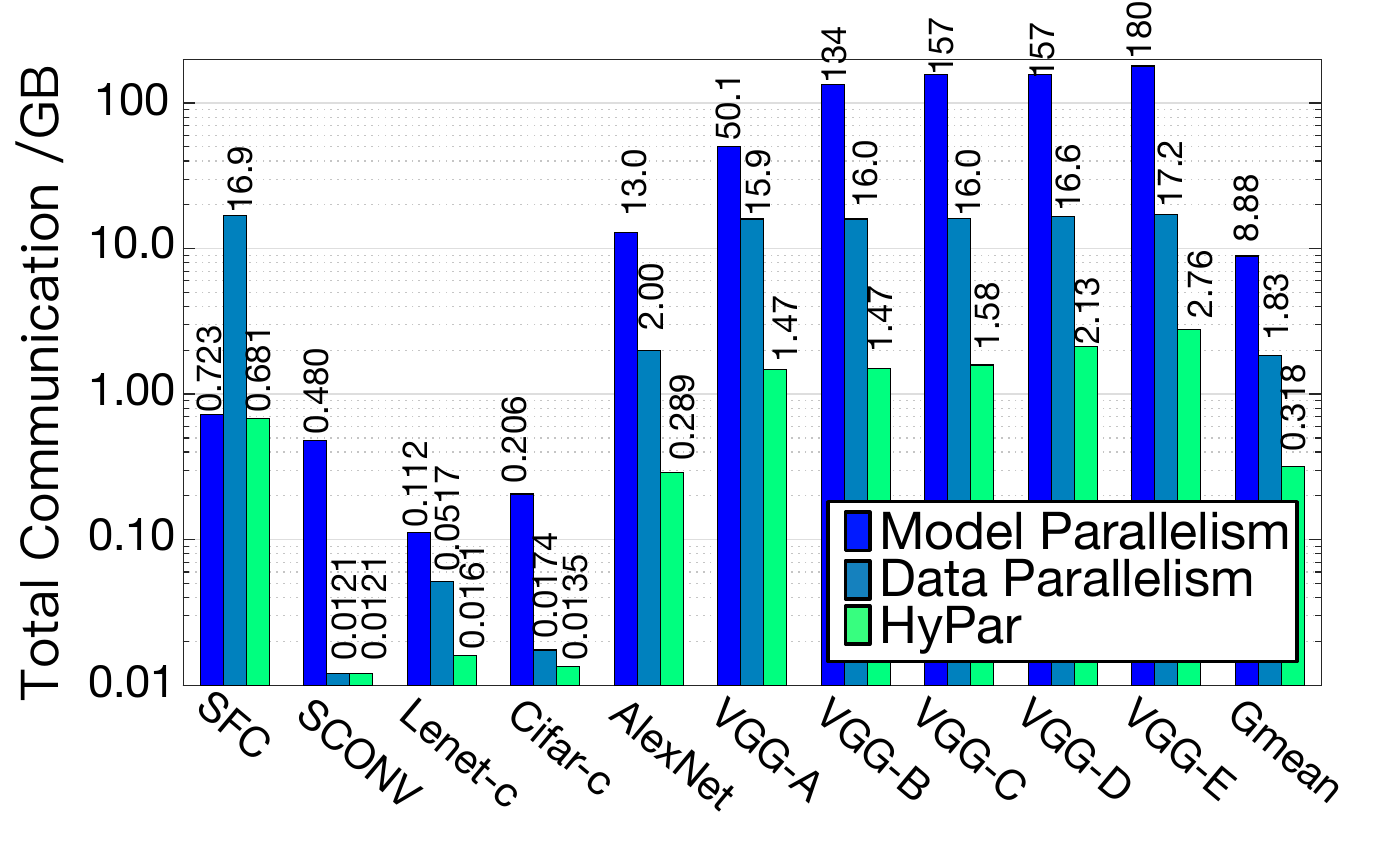}
\vspace{-9pt}
\caption{Total communication (in GB) of Model Parallelism, Data Parallelism and \hypar\ per step.}
\label{fig_eval_com}
\vspace{-0pt}
\end{figure}

In \hypar, the total communication of a network is optimized to 
improve the performance and energy efficiency. We show
the total communication per step of the default Model Parallelism, the default Data Parallelism and \hypar\ in Figure \ref{fig_eval_com}. 

The geometric means of total communication for Model Parallelism, Data Parallelism and \hypar\ are 8.88 GB, 1.83GB and 0.318 GB respectively.
Model Parallelism mostly has much higher amount of total communication than Data Parallelism and \hypar. However, for the extreme case {\tt SFC}, the total communication of Model Parallelism is lower than that of Data Parallelism. \hypar\ has lower amount of total communication (0.681 GB) than Model Parallelism (0.723 GB) in {\tt SFC}. For the other extreme case {\tt SCONV}, \hypar\ has the same amount of total communication as Data Parallelism, and the amount is lower than Model Parallelism. For other eight networks, especially the large size networks, i.e., {\tt AlexNet}, {\tt VGG-A}, {\tt VGG-B}, {\tt VGG-C}, {\tt VGG-D} and {\tt VGG-E}, the total communication in Data Parallelism is almost ten times lower than that of Model Parallelism, and \hypar\ is about another ten times lower than Data Parallelism. We can find that the communication is a key factor that determines the performance and energy efficiency of an array of accelerators.

\subsection{Case Studies}


\subsubsection{Parallelism Space Exploration for Lenet-c}

\begin{figure}[tb]
\vspace{-0pt}
\centering
\includegraphics[width=0.7\columnwidth]{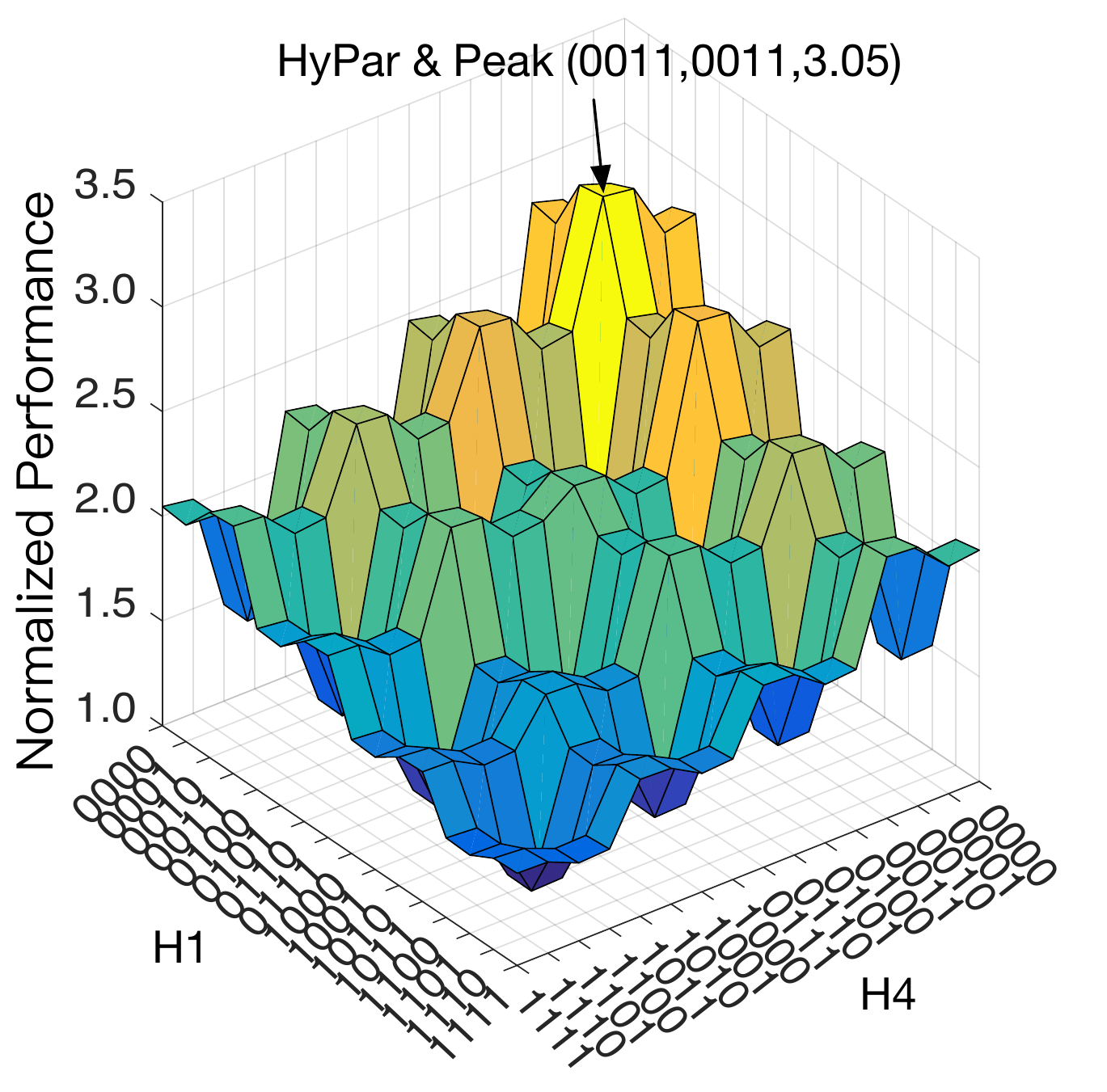}
\vspace{-9pt}
\caption{Normalized performance (to Data Parallelism) in parallelism space exploration for {\tt Lenet-c}. H2 and H3 are fixed and parallelism for layers at H1 and H4 are explored. 0 indicates data parallelism while 1 indicates model parallelism.}
\label{fig_eval_perf_space_lenet}
\vspace{-12pt}
\end{figure}

We explore the parallelism space for {\tt Lenet-c} to find the maximum performance. {\tt Lenet-c} has four weighted layers, and four partition hierarchy levels, so the capacity of searching space would be $2^{4\times 4} = 65536$, which is too large. As an alternative, we fix the parallelisms of all four layers at two hierarchy levels H2 and H3, and explore the possible parallelisms for all four layers at two hierarchy levels H1 and H4. The parallelisms of layers at H2 and H3 are fixed as the optimized ones, as shown in Figure \ref{fig_opt_par} (c). So the capacity of the searching space is now $2^{2\times 4} = 256$.

The results are shown in Figure \ref{fig_eval_perf_space_lenet}. The results are normalized to the default Data Parallelism. As we can see, the peak of the normalized performance is 3.05$\times$, at H1 = 0011 and H4 = 0011, which means the parallelisms for four layers at H1 are dp, dp, mp, mp and the four layers at H4 are dp, dp, mp, mp. That is exactly the performance gain of {\tt Lenet-c} with the parallelisms optimized by \hypar.

\subsubsection{Parallelism Space Exploration for VGG-A}

\begin{figure}[tb]
\vspace{-0pt}
\centering
\includegraphics[width=0.7\columnwidth]{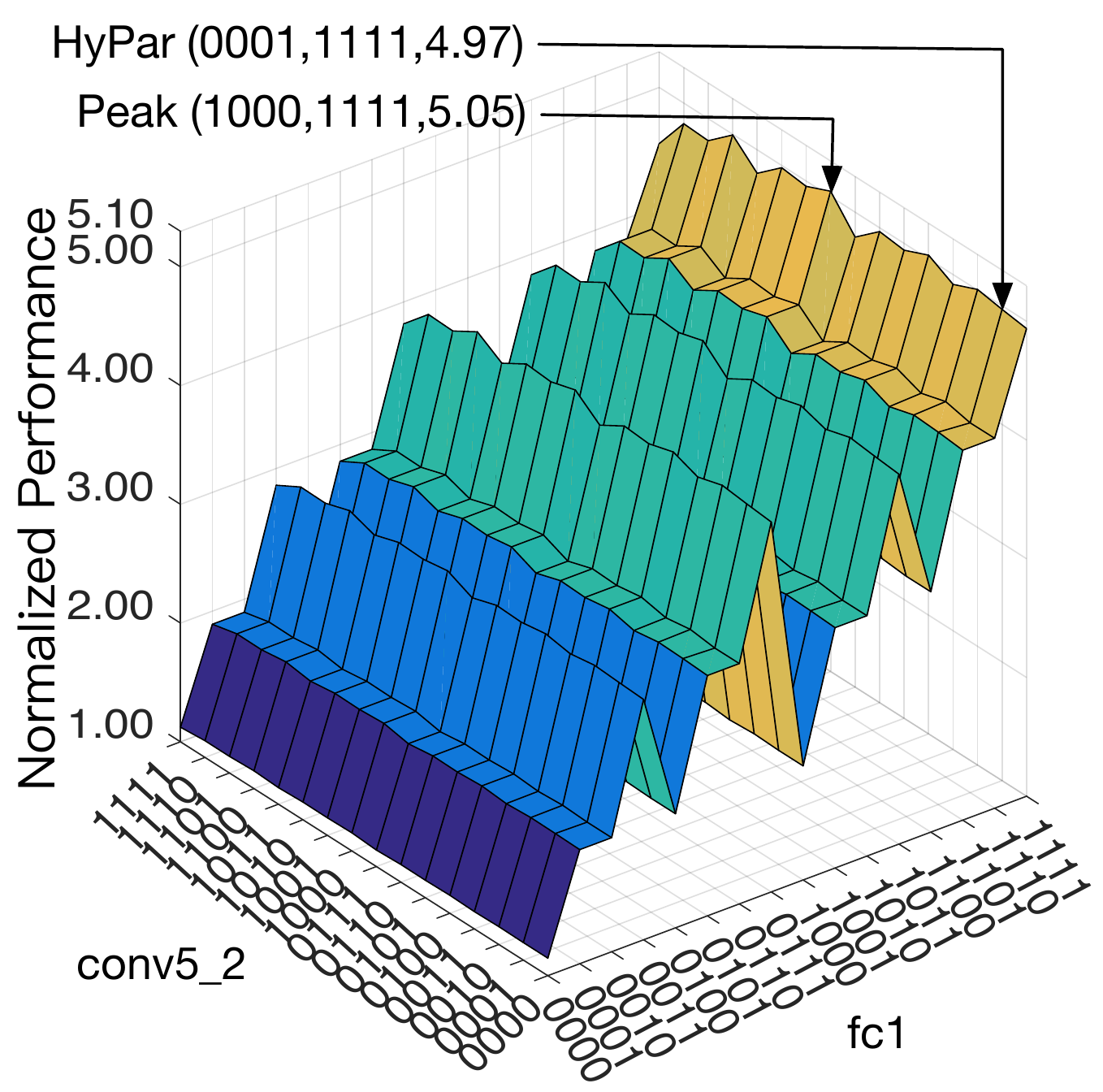}
\vspace{-9pt}
\caption{Normalized performance (to Data Parallelism) in parallelism space exploration for {\tt VGG-A}. All weighted layers are fixed except {\tt conv5\_2} and {\tt fc1}. Parallelism for H1 to H4 of {\tt conv5\_2} and {\tt fc1} are explored.}
\label{fig_eval_perf_space_vgga}
\vspace{-12pt}
\end{figure}

We explore the parallelism space for {\tt VGG-A} to find the maximum performance. While the capacity of full searching space would be $2^{4\times 11} = 17.6$ T, which is never possible to enumerate every point. As an alternative, we fix the parallelisms of nine layers in the network, except {\tt conv5\_2} and {\tt fc1}. We then explore the possible parallelisms for {\tt conv5\_2} and {\tt fc1} at the four hierarchy levels. The parallelisms of the other layers are fixed as the optimized ones, as shown in Figure \ref{fig_opt_par} (f). The capacity of the searching space is reduced to $2^{4\times 2} = 256$.

The exploration results are shown in Figure \ref{fig_eval_perf_space_vgga}. The results are normalized to the default Data Parallelism. As we can see, the peak of the normalized performance is 5.05$\times$, at {\tt conv5\_2} = 1000 and {\tt fc1} = 1111,
which means the parallelisms for {\tt conv5\_2} at four 
hierarchy levels (H1 to H4) are mp, dp, dp, dp and {\tt fc1}
in four hierarchy levels (H1 to H4)
are mp, mp, mp, mp. However, the performance optimized by \hypar\ is 4.97$\times$, and the corresponding parallelisms for {\tt conv5\_2} at four 
hierarchy levels (H1 to H4) are dp, dp, dp, mp. That is because \hypar\  optimizes the total communication as a proxy for 
optimizing performance. Even \hypar\ failed to provide the 
maximum performance with the optimized setting, the performance of \hypar\ is very close to the maximum (4.97$\times$ vs. 5.05$\times$), and is still much higher than the baseline Data Parallelism (4.97$\times$ vs. 1.00$\times$).

\begin{figure}[b!]
\vspace{-12pt}
\centering
\includegraphics[width=0.7\columnwidth]{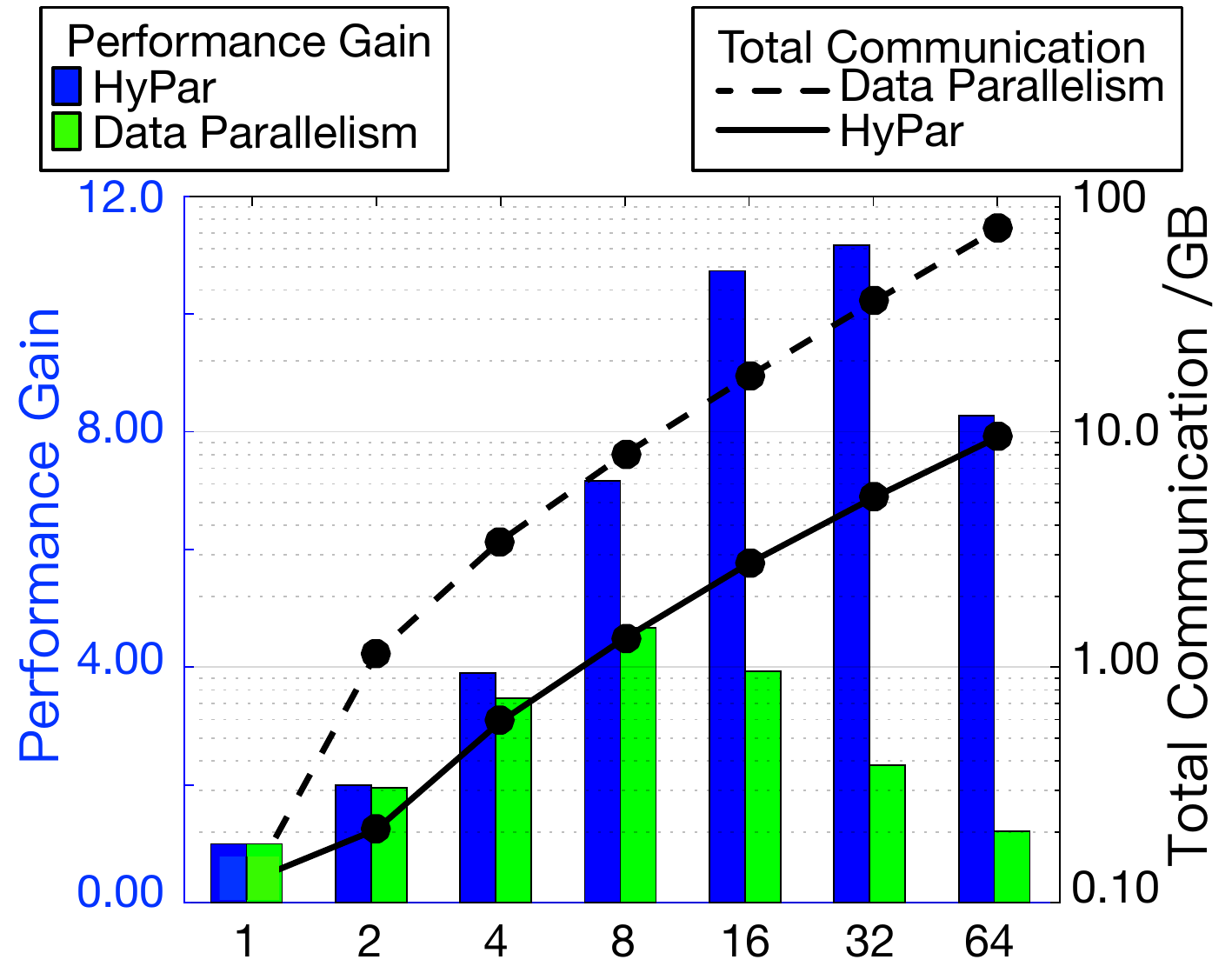}
\vspace{-9pt}
\caption{Comparison of scalability of \hypar\ and Data Parallelism. Left Y axis: performance gain normalized to one accelerator, right Y axis: total communication.}
\label{fig_eval_scal}
\vspace{-0pt}
\end{figure}

\subsection{Scalability}

We explore the scalability of \hypar\ using {\tt VGG-A} as an example. In the exploration, we scale the number of accelerators from 1 to 64. We compare the performance gains of \hypar\ and the default Data Parallelism. The gains are normalized to the performance of the performance of one accelerator.

The results are shown in Figure \ref{fig_eval_scal}. 
\hypar\ always performs better than the default Data Parallelism. The performance gains of the default Data Parallelism become decreasing after 
the number of accelerators exceeds 8. But the gains of \hypar\ increases until the number of accelerators 
exceeds 32. We can see that \hypar\ scales better than the default Data Parallelism. \hypar\ is always better than the default Data Parallelism in performance gains, and \hypar\ always has lower total communication.

\subsection{Comparisons}
\label{sec_eval_compare}

\subsubsection{Comparison of H Tree and Torus Topology}

We compare the performance of H tree and torus typologies. The parallelisms for each layers are the optimized
choices by \hypar, but the only difference is the connection topology of the sixteen accelerators. For H-tree, it is physically it is a fat-tree, and switches are placed at each parent node. The bandwidth between groups in a higher hierarchy are doubled compared to that of a lower hierarchy (but the number of links is halved). In a torus, the bandwidth for a link is the same. 

\begin{figure}[b]
\vspace{-12pt}
\centering
\includegraphics[width=0.95\columnwidth]{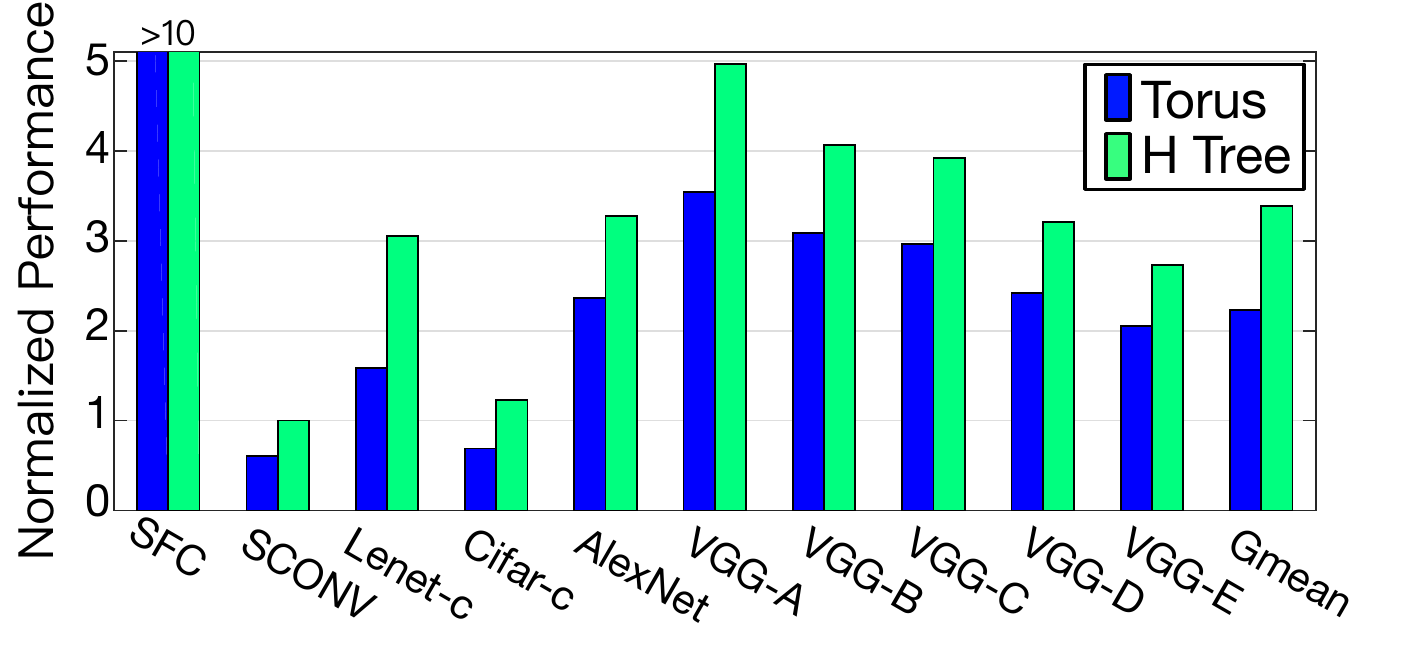}
\vspace{-9pt}
\caption{Normalized performance (to Data Parallelism) of torus and H tree topology.}
\label{fig_eval_toplology}
\vspace{-0pt}
\end{figure}

Figure \ref{fig_eval_toplology} shows the performance of torus and H tree topology, normalized to Data Parallelism. 
For {\tt SFC}, both the two typologies have a speedup
of more than $10\times$, because all layers in {\tt SFC} are 
fully-connected layers, and Data Parallelism has lower performance regardless of connection 
typologies. For the other networks, we can see, H tree outperforms torus topology, because the parallelism and 
tensor partition are determined in a binary tree pattern, and H tree is naturally more suitable for the pattern. The geometric mean of 
performance of torus and H tree typologies are 2.23$\times$ and 3.39$\times$.

While three are many different possible topologies for the accelerator array,
\hypar\ is topology independent. It is a simplification, but the topology-independent
communication model and the dynamic programming method indeed reduced the 
total communication between accelerators. From the comparison of H tree and torus, we can see 
the partition also works for torus
although \hypar\ prefers H tree, so the simplification is reasonable.

\subsubsection{Comparison of \hypar\ and the Trick in \cite{krizhevsky2014one}}
\label{sec_hypar_to_the_trick}
Batch size is an important hyper parameter 
in DNN training, and batch size 
should be customized for 
specific purposes rather than 
a default setting. For for larger 
training throughput \cite{goyal2017accurate}, 
a larger batch size (eg. 4096) 
should be selected, while for 
higher testing accuracy 
and ability to generalize \cite{masters2018revisiting}, 
a small batch size (eg. 32) 
should be selected.

\begin{figure}[tb]
\vspace{-0pt}
\centering
\includegraphics[width=0.7\columnwidth]{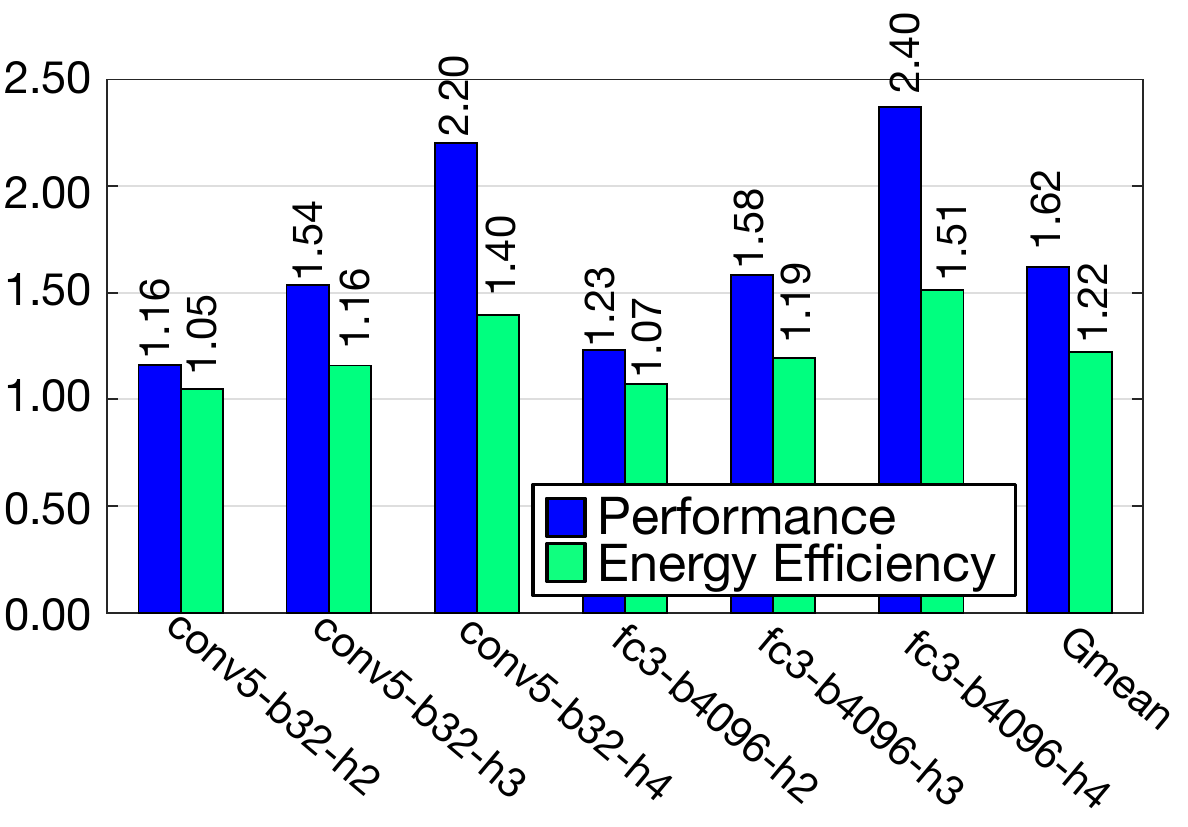}
\vspace{-9pt}
\caption{Performance and energy efficiency of \hypar\ compared to the trick in \cite{krizhevsky2014one}.}
\label{fig_eval_trick}
\vspace{-12pt}
\end{figure}

Thus we use the two batch size, 
i.e. {\tt b32} and {\tt b4096}, 
to evaluate \hypar\ and the Trick in \cite{krizhevsky2014one}. 
We use the fully-connected and convolutional layers {\tt fc3} 
and {\tt conv5} in {\tt VGG-E}, 
under three hierarchy levels {\tt h2}, {\tt h3} and {\tt h4}.
Figure \ref{fig_eval_trick} shows the performance and 
energy efficiency of \hypar\ 
compared to the Trick. We can see, the performance of \hypar\ 
is 1.62$\times$ better than 
the Trick and \hypar\ is 1.22$\times$ more energy efficient on average. \hypar\ can be 2.40$\times$ faster than the Trick.


So one may ask, what is wrong with the 
Trick \cite{krizhevsky2014one}? Let's get back to our communication model (Section \ref{sec_commodel}).
With the intra-layer communication model, for {\tt conv5}, 
$\mathbb{A}(\triangle\mathbf{W}_{l})=C_iC_oK^2=512\times 512\times 3^2=2,359,296$, while 
$\mathbb{A}(\mathbf{F}_{l+1})=BC_oWH=32\times 512\times 14\times 14=3,211,264$. 
Because $\mathbb{A}(\triangle\mathbf{W}_{l})$ $< \mathbb{A}(\mathbf{F}_{l+1})$,
here {\tt conv5} should be configured to model parallelism 
rather than data parallelism in the Trick. 
For {\tt fc3}, 
$\mathbb{A}(\triangle\mathbf{W}_{l})=C_iC_o=4096\times 1000=4,096,000$, while 
$\mathbb{A}(\mathbf{F}_{l+1})=BC_o=4096\times 1000=4,096,000$. 
$\mathbb{A}(\triangle\mathbf{W}_{l})$ and $\mathbb{A}(\mathbf{F}_{l+1})$ 
are the same, we can not clearly see 
which parallelism is better than 
the other, but we can further use 
inter-layer communication model to explain. 
According to Table \ref{table_inter_layer}, 
the communication of dp-dp is 0 
while that of either mp-mp or mp-dp is 
not 0. 
So we should choose data parallelism 
for that layer, but unfortunately, 
the Trick chose model parallelism. 

%% file: sec_conc.tex
We propose \hypar\ to determine layer-wise parallelism for deep neural 
network training with an array of DNN accelerators.
\hypar\ partitions the feature map tensors (input and output), the kernel tensor, 
the gradient tensor, and the error tensors for the DNN accelerators. 
The problem is to find the best parallelism
configuration for all weighted layers, 
a problem considered by several works~\cite{krizhevsky2014one,tofu,tofu2}.
The optimization target is to search a partition that
minimizes the total communication during training a complete deep neural network.
\hypar's dynamic programming and recursive partition based method, inspired by 
recent works ~\cite{tofu,tofu2}, is practical: the time complexity for the partition 
search in \hypar\ is {\em linear}.
We apply this method in \hypar\ architecture,
an HMC-based DNN training architecture to minimize data movement.
We evaluate \hypar\ with ten DNN models from classic Lenet to large-size model VGGs, and the number of weighted layers of these models ranges from four to nineteen. 
Our evaluation shows that \hypar\ achieves a performance gain of 3.39$\times$ and an energy efficiency gain of 1.51$\times$ compared to the default data parallelism on average, and \hypar\ performs up to 2.40$\times$ better than the trick \cite{krizhevsky2014one}.

\vspace{9pt}
\noindent\textbf{\large{ACKNOWLEDGEMENT}}
\vspace{0pt}

We thank the anonymous reviewers of HPCA 2019, MICRO 2018, ISCA 2018 for their constructive and insightful comments. This work is supported in part by NSF 1725456, 1615475, and DOE DE-SC0018064. This work is also supported by the National Science Foundation grants NSF-CCF-1657333, NSF- CCF-1717754, NSF-CNS-1717984, and NSF-CCF-1750656.